\documentclass{aastex62}

\graphicspath{{./}{figures/}}
\received{\today}
\revised{XXXX XX, 2019}
\accepted{XXXX XX, 2019}
\submitjournal{ApJ}

\shorttitle{}
\shortauthors{Lacy et al.}

\begin{document}

\title{Prospects for Directly Imaging Young Giant Planets at Optical Wavelengths}

\correspondingauthor{Brianna Lacy}
\email{blacy@astro.princeton.edu}

\author[0000-0002-0786-7307]{Brianna Lacy}
\affil{Princeton University \\
Peyton Hall, Ivy Lane \\
Princeton, NJ 08544, USA}

\author{Adam Burrows}
\affil{Princeton University \\
Peyton Hall, Ivy Lane \\
Princeton, NJ 08544, USA}

\begin{abstract}
In this work, we investigate the properties of young giant planet spectra in the optical and suggest that future space-based direct imaging missions should be considering young planets as a valuable and informative science case. While young planets are dimmer in the optical than in the infrared, they can still be brighter in the optical than a mature planet of similar mass. Therefore, an instrument designed to characterize mature planets should also be suitable for high-precision photometric imaging and spectroscopy of young self-luminous planets in a wavelength range and at a contrast ratio not attainable from the ground. We identify known young self-luminous companions which are feasible targets for WFIRST-CGI and compute spectra for them, including a treatment of scattering and reflected light at optical wavelengths. Using these results, we highlight potentially diagnostic spectral features that will be present in the WFIRST-CGI wavelengths. Expanding to direct imaging missions beyond WFIRST-CGI, we also use evolutionary models across a grid of masses and planet-star separations as inputs to compute spectra of hypothetical objects, exploring when reflected light may contribute to a degree comparable to that of thermal emission from the residual heat of formation.
\end{abstract}


\section{Introduction} \label{sec:intro}
Current ground-based direct imaging of self-luminous planets is mostly done at near infrared wavelengths (NIR), where Strehl ratios and planet-star contrast ratios are favorable (\citealt{GPI}; \citealt{CHARIS}; \citealt{SCExAO}; \citealt{SPHERE}; \citealt{ALES}). The coverage of most instruments only reaches as short as 0.8$\mu$m. There are a few exceptions to this trend, wherein planets are observed directly at optical wavelengths. For example, some groups have targeted H-$\alpha$ emission, thought to be related to accretion onto protoplanets (\citealt{sallum2015}; \citealt{muller2018}), but these observations are limited to the youngest systems still undergoing active accretion. SPHERE-ZIMPOL on the VLT, a high-resolution polarization camera which covers the optical range from 0.6-0.9 $\mu$m \citep{schmid2018}, has been used to characterize disks and place upper limits on the levels of polarization from known planets. Outside of the polarization mode it only attains contrast ratios on the order of 10$^{-3}$. The Hubble Space Telescope's Space Telescope Imaging Spectrograph (HST-STIS) has been used in a coronagraphic mode to observe circumstellar disks and to place upper limits on the optical brightness of some known planets (\citealt{Grady2003}; \citealt{Apai2015}; \citealt{Debes2019}). It can reach contrast ratios of 10$^{-6}$ which is sufficient for observing brown dwarf companions to main sequence stars from $\sim$0.1-1.0 $\mu$m and possibly for observing the hottest planetary mass companions from $\sim$0.8-1.0 $\mu$m.

In the future, space-based high-contrast imaging facilities will be able to obtain spectra of planets at higher contrasts, bluer wavelengths, and smaller working angles than what is currently possible. The coronagraphic instrument on the Wide Field Infrared Survey Telescope (WFIRST-CGI, \citealt{Spergel2013}) is a technology demonstrator set to fly in 2025 and intended to pave the way for direct imaging with a great observatory in the 2030s (\citealt{habex2019}; \citealt{luvoir2019}). WFIRST-CGI mission requirements are set to achieve contrast ratios around $\sim$5$\times$10$^{-8}$, at wavelengths of 0.55 $\mu$m to 0.85 $\mu$m, and separations of 0.15$''$ to 1.5$''$. Current predicted performance by team engineers puts the contrast ratios at 10$^{-9}$. Designs for later great observatories aim for a better contrast ratio of $\sim$10$^{-10}$, broader wavelength coverage from 0.2 to 2.0 $\mu$m, and to reach smaller separations of 0.05$''$ to 0.33$''$. The planned capabilities of these missions are largely motivated by the goal of characterizing earth-size planets in the habitable zones of nearby G stars, but such missions will also provide exquisite data for many other types of exoplanets. Capabilities from the ground will also reach higher contrasts and shorter wavelengths with the advent of Extremely Large Telescopes (ELT). While they won't be included for first light operations, eventually ELTs could implement dedicated optical wavelength AO systems, and achieve contrast ratios on the order of 10$^{-7}$ as short as 0.6 $\mu$m \citep{Bowler2019}. 

Motivated by this observational context, we present a theoretical consideration of what might be seen when we observe young giant planets at optical wavelengths. We demonstrate that young giant planets provide an intriguing additional science case that should be included in planning and advocating for future direct-imaging missions with optical capabilities. Optical spectra of young planets will be bright enough for high signal-to-noise observations. Combining optical and NIR wavelength coverage will provide the best chance of unambiguously characterizing a planet's effective temperature, surface gravity, atmospheric metallicity, and condensate properties. Furthermore, obtaining both NIR and optical spectra presents a unique opportunity to empirically measure the geometric albedo and phase function of reflected light on an exoplanet at orbital periods greater than a few days.

There are still a lot of unanswered questions surrounding young giant planets. A few dozen self-luminous planets and substellar companions have been characterized with ground-based direct imaging so far (see \citealt{Bowler2016} for a recent review). Characterization efforts generally compare NIR spectra and photometry with a variety of available models (e.g. \citealt{madhu2011}; \citealt{Marley2012}; \citealt{Currie2013}; \citealt{Currie2014}; \citealt{Baraffe2015}) and template libraries of brown dwarf spectral types. These studies have provided estimates of temperatures and surface gravities for most of the directly-imaged substellar companions, but credible measurements of metallicity and C/O ratios will require the capabilities of future instruments (\citealt{biller2018}). Extending wavelength coverage into the optical would help with this effort. Metallicity measurements are valuable because they can inform our understanding of planet formation, especially if we can obtain them for a sizeable population of giant planets and substellar companions. 

One intriguing trend emerging from observations so far, is that young giant planets and low surface gravity brown dwarfs extend the L dwarf sequence to redder colors and fainter absolute magnitudes than their more massive brown dwarf cousins, implying a delayed transition from cloudy atmospheres to condensate-free T dwarfs (\citealt{Ackerman2001};\citealt{Chauvin2004}; \citealt{Burrows2006}; \citealt{Metchev2006}; \citealt{Marois2008}; \citealt{Bowler2010}; \citealt{Patience2010}; \citealt{Bowler2013};  \citealt{Faherty2013}; \citealt{Liu2013}; \citealt{Filipazzo2015}). This trend, along with evidence for non-equilibrium chemistry, is commonly attributed to changes in the details of cloud microphysics and atmospheric mixing at lower surface gravities (\citealt{Burrows2006}; \citealt{Hubeny2007}; \citealt{Barman2011}; \citealt{Marley2012}; \citealt{Ingraham2014}; \citealt{Skemer2014}; \citealt{Zahnle2014}; \citealt{Miles2018}), but the details are not fully understood. Observing young giant planets at optical wavelengths could provide constraints on the nature of condensates in young giant planets and perhaps help to uncover the mechanisms delaying the L-T transition in low-surface-gravity atmospheres.

Extending observations to optical wavelengths opens up the possibility of detecting reflected light off of young giant planets. Models used to characterize self-luminous planets so far have not considered the effects of optical stellar irradiation and reflection, because the objects observed to date are on very wide orbits, and are observed at NIR wavelengths, where reflected light's contribution is negligible. The only measurements made of exoplanet optical albedos have come from phase curves of hot Jupiters \citep{placek2016}. Optical phase curves are available from Kepler and K2 (\citealt{esteves2015}), and many more will be collected by TESS \citep{mayorga2019}. Geometric albedos for hot Jupiters have been observed to be quite low ($\leq$0.1, \citealt{Burrows2007}; \citealt{heng2013}; \citealt{mallon2019}). \cite{parmentier2018} provide a recent review of the theoretical expectations and observational results for planetary phase curve studies of hot Jupiters, emphasizing the qualitative agreement, but quantitative disagreement between theory and observation. 

Comparing the albedo properties of hot Jupiters and a population of directly-imaged moderate-separation young giant planets would prove an interesting exercise in comparative planetology, and perhaps help to inform some of the questions left open by phase curve observations so far. \cite{sudarsky2000} suggest albedos of mature giant planets ought to separate into classes of objects based primarily upon equilibrium temperature and the properties of the condensates which form at depth or at altitude in different temperature ranges. There is an overlap in equilibrium temperatures for younger and more massive wide-orbit objects, but Hot Jupiters typically exhibit steep day-night temperature gradients compared to the what is expected for the wider orbit objects we consider in this work. This will result in different atmospheric dynamics, vertical mixing, and cold trapping between the two populations, despite similar average temperatures. 

Observing young giant planets at optical wavelengths poses a more compelling science case for future missions if we can expect to observe a reasonable population of them. Recent occurrence rate estimates from radial velocity (RV) and direct imaging surveys combined indicate that giant planet occurrence rates increase outwards from the ultra-short period Hot Jupiters, peak between 1 and 10 AU and then decrease out towards 100 AU (\citealt{Baron2019}; \citealt{Fernandes2019}; \citealt{Nielsen2019}; \citealt{Wagner2019}). These studies find occurrence rates in the neighborhood of 3.5\% for 5-13 M$_J$ planets between 10 and 100 AU for all stars. Existing coronagraphs generally have access to planets residing between 10 and 1000 AU from their host stars, and attain contrast ratios sensitive to planet masses of around 2 M$_J$ at the smallest \citep{Nielsen2019}. This leads us to expect an abundance of slightly older or smaller planets hiding below the best contrasts achieved so far, and a large number of young planets hidden interior to the working angles of current instruments. GAIA astrometry alone is expected to find many of these unseen giant planets, yielding around 20,000 new 1-15 M$_J$ planets out to $\sim$5 AU separations \citep{Perryman2014}. 

This paper is organized as follows: In \S \ref{sec:target_selection}, we identify known systems with self-luminous substellar companions that would be possible targets for the WFIRST-CGI technology demonstrator. We also look at occurrence rates from RV and ground-based direct imaging surveys, along with a sample of nearby young stars, to estimate how many currently undetected young planets might be observable in the future. In \S \ref{sec:models}, we describe our modeling process. In \S \ref{sec:known}, we present theoretical spectra for known young giant planets using planet-star characteristics drawn from the literature. We use these spectra to point out diagnostic features that fall within the optical wavelength range, and simulate WFIRST-CGI observations. In \S \ref{sec:hypothetical}, we explore hypothetical systems, looking at how spectra evolve over time for a variety of companion masses and star-companion separations. We consider the effects of adding condensates to the atmospheres, and explore when reflected light contributes to spectra as a function of mass and planet-star separation. We conclude with a summary of key results in \S \ref{sec:conclusion}. 

\section{Candidate Targets for WFIRST-CGI}\label{sec:target_selection}
The WFIRST mission is intended to fly in 2025 and will include a technology demonstration of an optical coronagraph  \citep{Spergel2013}, applying techniques like wave-front sensing and control for the first time in space. As part of its technology demonstration, WFIRST-CGI will obtain images and low resolution spectra of a few mature giant planets and a few young giant planets. It will also characterize exozodiacal dust and obtain polarimetry maps of circumstellar disks at unprecedented sensitivity \citep{2019summary}. If all goes well in the technology demonstration, then there will be an extended mission phase offering opportunities for more observations.

The instrument consists of two coronagraph architectures paired with an EMCCD and a slit-prism spectrograph which will achieve an average spectral resolution of R$\sim$50 (\citealt{emccd}; \citealt{actualMissionReqs}; \citealt{spieWFIRST}). There will be $\sim$10\% width imaging filters centered at 0.575 and 0.825 $\mu$m, and two adjacent spectral regions with widths of $\sim$15\% centered on 0.66 and 0.73 $\mu$m. Table \ref{wfirst_tab} summarizes the WFIRST-CGI capabilities planned as of this writing. The details are likely to change slightly as the mission moves to final design stages. Updated values can be accessed via the IPAC website\footnote{https://wfirst.ipac.caltech.edu/sims/Param\_db.html\#coronagraph\_mode}. Along with many of the subsequent spectra shown in this work we provide indications of WFIRST-CGI's capabilities for easy reference; these include colored rectangles marking the wavelength range of each filter, and black horizontal lines denoting the baseline requirement contrast and the predicted contrast. 

\begin{table}[htb!]
    \centering
    \begin{tabular}{c|c|c|c|c|l|l}
Coronagraph & Dark Hole         & Working & Baseline  & Predicted  & Bands & Intended Use \\
         Type        & Shape             & Angles & Contrast  & Contrast   &         & \\
\hline
Shaped Pupil    & Bowtie, 130$^{\circ}$&  3-9 $\frac{\lambda}{D}$ & 5$\times$10$^{-8}-$10$^{-7}$ & 6$\times$10$^{-10}$ &2 (0.66 $\mu$m) & Reflected light spectra,\\
& &   &  &   &3 (0.73 $\mu$m)  &young giant planet spectra\\
Shaped Pupil   & Disk, 360$^{\circ}$ &6.5-20 $\frac{\lambda}{D}$ & 5$\times$10$^{-8}-$10$^{-7}$ & 2$\times$10$^{-9}$ & 4 (0.825 $\mu$m)  & Disk imaging,  \\
 & &   &  & &  &disk polarimetry\\
Hybrid-Lyot  & Disk, 360$^{\circ}$  & 3-9 $\frac{\lambda}{D}$   & 5$\times$10$^{-8}-$10$^{-7}$ & 6$\times$10$^{-10}$ &1 (0.575 $\mu$m) & Reflected light images, \\
 & &  &  &  &   & planet search\\

    \end{tabular}
    \caption{Summary of WFIRST-CGI baseline requirement capabilities. Note that the quoted values are for a star with M$_{V*}\leq$ 5. Updated capabilities can be found at https://wfirst.ipac.caltech.edu/sims/Param\_db.html\#coronagraph\_mode.}
    \label{wfirst_tab}
\end{table}

To identify potential WFIRST-CGI targets, we collected a list of all the confirmed directly-imaged substellar companions included in the exoplanet encyclopedia\footnote{http://exoplanet.eu} and the NASA exoplanet archive\footnote{https://exoplanetarchive.ipac.caltech.edu}. We then applied two cuts. First we selected companions with a host star M$_{V*}$ $<$ 8. We then checked which of these planets fall in the working angles of WFIRST-CGI's two ranges of 3-9 $\frac{\lambda}{D}$ for $\lambda$=0.575-0.73 $\mu$m, and 6.5-20 $\frac{\lambda}{D}$ for $\lambda$=0.784-0.866 $\mu$m. This search resulted in five companions which fall within the $\lambda$=0.546-0.785 $\mu$m coverage (imaging filter 1 and spectral bands 2 and 3), and nine objects which fall within the $\lambda$=0.784-0.866 $\mu$m coverage (imaging filter 4). 51 Eri b and $\beta$-Pic b fall fully within both observing modes at some point in their orbits. We then searched through the literature for estimates of the properties relevant to modeling the companions' spectra: effective temperature, radius, surface gravity, host star type, and planet-star separation. These are listed in Table \ref{ygp_prop_tab}. It should be noted that some of these values have a large uncertainty, but for our purposes this is not critically important. We aim for an order of magnitude sense of whether observing young giant planets is feasible at optical wavelengths, and a general idea of what spectral features are expected. From this group, we chose the five fully-formed objects accessible in the shorter wavelength range for further spectral modeling and exposure time calculations (marked with a * in the table).   

\begin{table}[htb!]
    \centering
    \begin{tabular}{c|c|c|c|c|c|c|c|c|c|l}
Object & Stellar &Stellar& Distance & Age & Semi-major & Mass & Radius &  T$_{eff}$&log(g)& Reference\\
Name& Type&M$_V$&(pc)&  (Myr) &axis (AU) &(M$_J$)& (R$_J$)&(K)&(dex)&\\
\hline
51 Eri b* & F0IV &5.223 & 29.4 & 20& 12 & 2.6 & 1.11 &  700&3.5&\citealt{eri-b}\\
HD 95086b & A8III &7.36  & 90.4 &17  & 61.7 & 5 & 1.3 &1050  & 3.85 &\citealt{DeRosa2016}\\
HIP 65426b &A2V&7.01&109.2&14&92&9&1.5&1450&4.5&\citealt{Chauvin2017}\\
HR 8799c & A5V & 5.96 & 39.4  & 60 &  42.9 &  10 &  1.35 &  900 & 4 &\citealt{Marley2012}\\
HR 8799d & A5V & 5.96 & 39.4 & 60 & 27 & 8.3 & 1.35 &  1000 &4&\citealt{Marley2012}\\
HR 8799e* & A5V & 5.96 & 39.4 & 60 & 16.4 & 9.2 & 1.17 &  1150 &4 &\citealt{Marois2010}\\
PDS 70c* & K5  &12 & 113.43 & 5.4& 20.6  & 10  & 1.6 &  1550& 4&\citealt{Christiaens2019_b}\\
$\beta$-Pic b* & A6V &3.86 & 19.3 & 40& 11.8 & 12.9 & 1.65 & 1750&4&\citealt{Snellen2018}\\
$\kappa$-And b &B9V&4.14&51.6&47&$\sim$75&13&1.2&1850&4.5&\citealt{Currie2018} \\
HD 206893B* & F5V &6.67 & 38.34 & 1100& 10 & 22.5 & 1.1 & 1400&4.75&\citealt{Grandjean2019}\\
HR 2562B & F5V &6.114 & 33.63 & 600& 20.3 & 26 & 1.11 & 1200&4.7&\citealt{Konopacky2016}\\
HR 3549B   &A0V&6.01&92.5&125&80&45&1.13&2350&4.9&\citealt{Mesa2016}\\
HD 984B* & F7V &7.3 & 47.1 & 80 & 18 & 48 & 1.26&  2730 &  &\citealt{Johnson-Groh2017}\\
HD 1160B   &A0V&7.12&103.5&50&80&79&&3000&&\citealt{Maire2016}\\
    \end{tabular}
    \caption{A selection of known self-luminous planets, brown dwarf companions, and accreting protoplanets that might be observable with WFIRST-CGI. Objects marked with a * will be visible with the full wavelength coverage of the 0.575-$\mu$m imaging band and the two spectral bands. Theoretical spectra for these objects, aside from PDS 70 c, are shown in \S \ref{sec:known}. Other objects will be visible in the longer wavelength 0.825-$\mu$m imaging band, and in some cases the longest wavelengths of the 0.76-$\mu$m spectral band.}
    \label{ygp_prop_tab}
\end{table}

In addition to planets imaged at NIR wavelengths, accreting protoplanets detected via H-$\alpha$ emission are possible targets for WFIRST-CGI. One of the engineering filters for WFIRST-CGI is likely to be centered on the H-$\alpha$ feature around 0.656 $\mu$m, and spectral band 2 spans $\lambda$=0.61-0.71 $\mu$m. Out of the various candidate protoplanets, PDS 70 b and c have emerged as robust detections (\citealt{keppler2018}; \citealt{haffert2019}) and, fortunately, both happen to fall within or near the working angles of WFIRST-CGI.  A big drawback to the PDS 70 system is that it is at a distance of 113 pc, so the visual magnitude of the host star is only 12. This is well below the brightness necessary for the WFIRST-CGI wavefront control systems to operate with optimal performance, so further study beyond the scope of this paper is necessary to check whether the system is observable. Nonetheless, obtaining  high signal-to-noise ratio optical spectra would certainly aid in the characterization of these objects and any similar systems found in the future. Observations of H-$\alpha$ emission from both b and c have been done (\citealt{wagner2018}; \citealt{haffert2019}). PDS 70 b also has spectra and imaging spanning 1.0 $\mu$m to 4.0 $\mu$m (\citealt{muller2018}; \citealt{Christiaens2019_a}). \cite{Christiaens2019_b} analyzed all these observations together and concluded that they are best fit by a model which includes a circumplanetary disk.

\begin{figure}[htb!]
\includegraphics[width=0.55\textwidth]{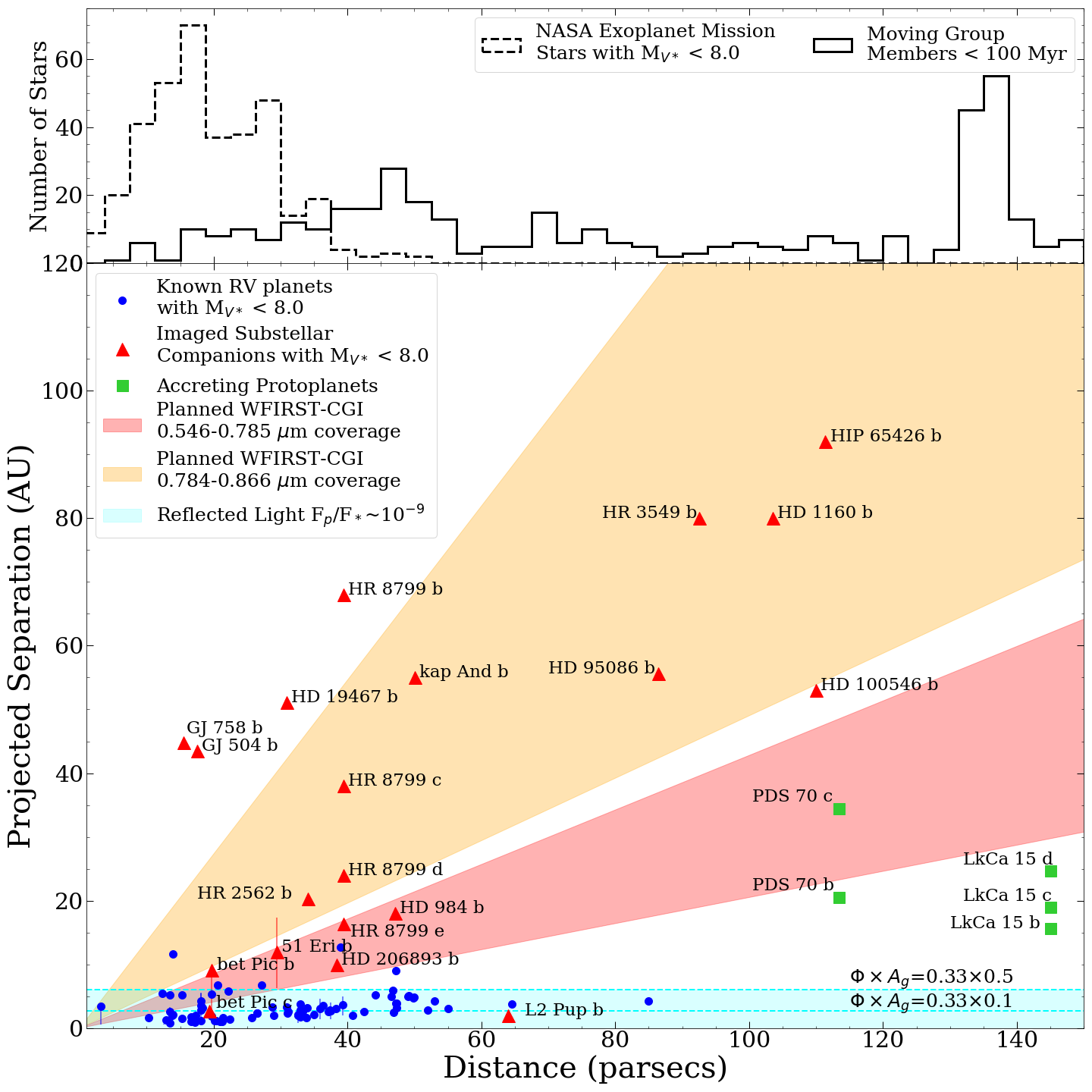}
\caption{\label{identify_targets} A visualization of WFIRST-CGI's working angles relative to the landscape of known and potential targets (recall that the angular separation between a planet and its host star is the ratio of the projected planet-star separation over the distance to the system). Shaded wedges indicate combinations of projected planet-star separation and system distance which will fall within the full wavelength coverage of the  SPC-bowtie configuration and HLC (red, filters 1,2,3 covering 0.546-0.785 $\mu$m) and the SPC-disk configuration (orange, filter 4, 0.784-0.866 $\mu$m). The variation of projected planet-star separation over the course of an orbit is indicated with a vertical line for systems with reasonably well-known orbits. The cyan blue shaded region indicates the projected separations at which a one Jupiter-radius planet would achieve contrast ratios of 10$^{-9}$ with reflected light alone, assuming a circular orbit. This is dependent on the phase of observation and the geometric albedo, as indicated by dashed lines. The histograms above the main figure are intended to convey a sense of how many potential stars could be searched for as-yet undetected planets at different distances. The dashed black line is the catalog of stars identified by NASA as targets for a future exoearth search. The solid black line is a collection of moving group members younger than 100 Myr.}
\end{figure}

Figure \ref{identify_targets} shows the plane of distance and projected planet-star separation with directly imaged self-luminous planets, radial-velocity planets, and accreting protoplanets plotted atop shaded wedges denoting the working angles of WFIRST-CGI. Recall that the working angle is simply the ratio of the projected planet-star separation over the distance to the system, so any target which lies within a shaded wedge will be within the working angles of the coronagraph. For reference, we also shade the projected separations within which a mature exo-Jupiter on a circular orbit would have a flux ratio of $\geq$10$^{-9}$ due to reflected light (cyan) assuming geometric albedos (A$_g$) and phase functions ($\Phi$) as indicated in the figure. A contrast floor of 10$^{-9}$ was the original target capability of WFIRST-CGI, and remains the goal for the engineering team despite the less stringent requirement for technology demonstration success. A number of studies have explored the prospects for characterizing mature planets with both the originally planned capabilities of WFIRST-CGI (\citealt{cahoy2010}; \citealt{Lupu2016}; \citealt{Robinson2016}; \citealt{Nayak2017}) and scaled back versions more in keeping with the current technology demonstrator status (\citealt{Batalha2019}; \citealt{Lacy2019}). 

Two histograms are shown across the top panel of Fig. \ref{identify_targets}, intended to provide a sense of how many as-yet unknown planets might possibly be observable with WFIRST-CGI. An extensive literature on the detectability of mature giant planets and predicting yields of blind searches is already available from works planning future hunts for exoearths (e.g. \citealt{exozodi}). \cite{savransky2010}, \cite{greco2015}, and \cite{savransky2016} consider prospects for a blind search with the capabilities of WFIRST-CGI in mind. Studies like these have informed the selection of stars for the NASA Exoplanet Mission Stars Catalog, shown by the dashed black line histogram. We also include a histogram of stars with ages less than 100 Myr (solid black line). These are the stars which provide a chance of observing young planets at a given separation. The 400 young stars in this sample have age estimates based upon their membership in co-moving groups (\citealt{torres2008}; \citealt{Zuckerman2004}). 

\begin{figure}
    \centering
    \includegraphics[width=0.85\textwidth]{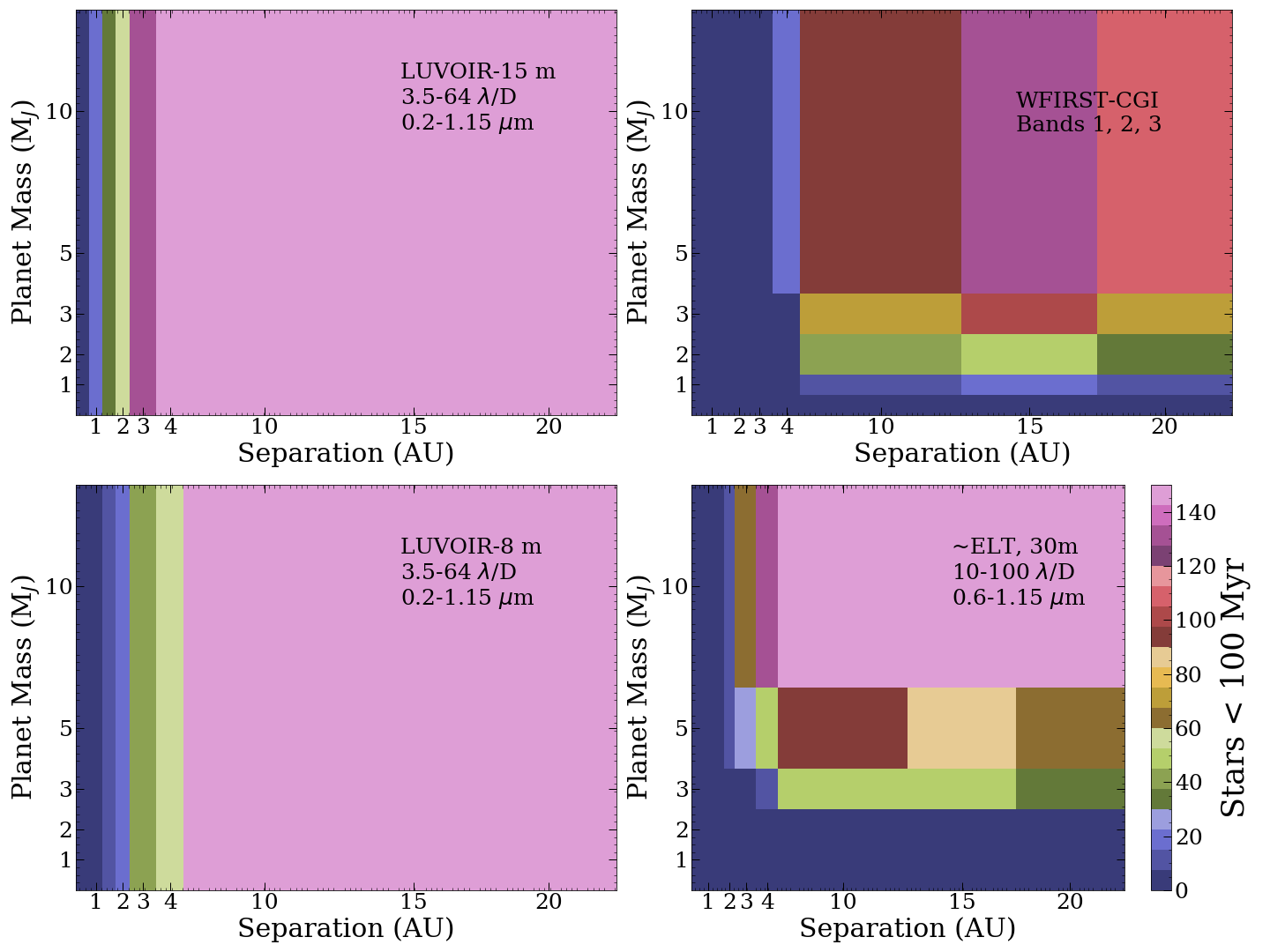}
    \caption{Visualization of how many nearby young stars reside at a distance such that we could observe planets with different combinations of mass and separation in the optical. Each combination of separation and planet mass is given a color indicating how many stars have an age and distance such that this planet would be observable within the working angles and contrast capabilities of the specified direct imaging mission. The top left panel shows the 15-m version of LUVOIR, the top right panel shows the 0.546-0.785 $\mu$m wavelength coverage of WFIRST-CGI, the bottom left panel shows the 8-m version of LUVOIR, and the bottom right panel shows an approximate ELT with second-generation optical AO. We consider a sample of 400 young stars within 150 pc thought to be part of moving groups. Note that, in order to better resolve the behavior at the smaller masses and separations, we set the colorbar maximum to 150 stars rather than 400. We use scaled bolometric luminosities from the grid of evolutionary models described in \S \ref{sec:hypothetical} to estimate whether a planet at a given mass and separation might be detectable in the optical. This procedure ignores variation in the stellar type of the young stars and does not consider the limit of a visual magnitude brighter than 8 used in Fig. \ref{identify_targets} for WFIRST-CGI.}
    \label{fig:tongue}
\end{figure}

 Moving beyond known targets and WFIRST-CGI mission planning, in the future we expect some new and interesting self-luminous planets to be observable at optical wavelengths. We provide another way to visualize potential targets in Fig. \ref{fig:tongue}. The color of each square indicates the number of young stars around which a planet with that particular combination of orbital separation and mass would be observable. The irregular binning in separation and mass correspond to the grid of evolutionary models we present later in the paper (see \S \ref{sec:hypothetical}). In this case we consider not only whether planets are within the coronagraph working angles, but also the limiting mass visible based on the estimated stellar ages. As we will discuss later in this work (see \S \ref{sec:disentangle}), to directly observe a planet with reflected light dominating at wavelengths less than 0.6 $\mu$m and self-luminosity dominating at wavelengths greater than 0.8 $\mu$m, then one must look for planets that are both young and at separations of a few AU or less. The panels on the left of Fig. \ref{fig:tongue} show that a future flagship mission in the style of LUVOIR should be able to observe at least a few of these exciting planets, considering the sample of young stars accessible and the expected occurrence rates of giant planets is $>$1\%. ELTs equipped with optical wavelength AO could also observe close-in planets, if they have masses above 3 M$_J$.  Unfortunately, WFIRST-CGI only reaches a substantive sample of over fifty stars once you limit to planets $>$2 M$_J$ in mass and at separations of 10 AU and outwards. This is not yet probing the regime of peak giant planet occurrence rate in mass or separation. WFIRST-CGI's inner working angle is not significantly smaller than current ground-based facilities, but it will reach down one to three orders of magnitude deeper in contrast. Systems with known directly imaged planets are a great place to look for additional closer-in planets as working angles improve, or slightly dimmer planets as contrast ratios improve. A recent example of this is the $\beta$-Pic system. \cite{Lagrange2019} detected an additional planet, $\beta$-Pic c, with a semi-major axis of only 2.69 AU. 

\section{Model Description} \label{sec:models}
We generated spectra with reflection using the atmosphere and spectral code CoolTLusty described and applied in \cite{Hubeny1995}, \cite{Hubeny1988},  \cite{Burrows2006} and \cite{Burrows2008}. CoolTLusty solves self-consistent hydrostatic atmospheres under stellar irradiation. It assumes chemical equilibrium with an ad hoc rainout scheme, employing a detailed pre-tabulated suite of thermochemical and opacity data (described in \citealt{Burrows1999} and \citealt{Sharp2007}, with updated measurements and linelists drawing from EXOMOL: \citealt{Yurchenko2014}). As input, this code requires the stellar spectral type for incoming irradiation, the planet-star separation, the planet radius, the planet surface gravity, and the planet effective temperature. To compute spectra for known self-luminous planets (\S \ref{sec:known}), we use the values listed in Table \ref{ygp_prop_tab}, and generally assume a condensate-free, solar metallicity atmosphere. 

To compute spectra for hypothetical planets (\S \ref{sec:hypothetical}) we first compute an irregular grid of evolutionary models using methods outlined in \cite{Burrows1997}, \cite{Burrows2001}, and \cite{Spiegel2012}. Our grid includes masses of 0.5, 1.0, 2.0, 3.0, 5.0, and 10.0 Jupiter-masses, and separations of 0.5, 1.0, 1.5, 2.0, 3.0, 4.0, 10.0, 15.0, and 20.0 AU. We always use a G2V type star for the stellar irradiation and assume solar metallicity. Atmospheres within the evolutionary models are assumed to have zero Bond albedo. We chose a subset of our evolutionary model grid for which to compute spectra. The input parameters for this subset are listed in Table \ref{hyp_tab}. Unless otherwise specified, when computing spectra, we again use stellar irradiation from a G2V type star and assume a clear atmosphere with solar metallicity. 

\begin{table}[htb!]
    \centering
    \begin{tabular}{c|c|c|c|c|c}
Planet Mass & Semi-major axis  & Age &  Planet Radius  & T$_{eff}$ &Surface Gravity\\
M$_J$) & (AU)& (Myr) & (R$_J$) & (K) & log(g cm/s$^2$)\\
\hline
1.0  &  1  &  1.00  &  1.690  &  803  &  2.94 \\
1.0  &  1  &  4.64  &  1.481  &  617  &  3.05 \\
1.0  &  1  &  21.54  &  1.322  &  443  &  3.15 \\
1.0  &  1  &  100.00  &  1.233  &  318  &  3.21 \\
1.0  &  10  &  1.00  &  1.690  &  803  &  2.94 \\
1.0  &  10  &  4.64  &  1.478  &  614  &  3.06 \\
1.0  &  10  &  21.54  &  1.313  &  431  &  3.16 \\
1.0  &  10  &  100.00  &  1.209  &  283  &  3.23 \\
2.0  &  1  &  1.00  &  1.593  &  1106  &  3.29 \\
2.0  &  1  &  4.64  &  1.414  &  856  &  3.39 \\
2.0  &  1  &  21.54  &  1.287  &  572  &  3.48 \\
2.0  &  1  &  100.00  &  1.211  &  388  &  3.53 \\
2.0  &  10  &  1.00  &  1.593  &  1106  &  3.29 \\
2.0  &  10  &  4.64  &  1.414  &  856  &  3.39 \\ 
2.0  &  10  &  21.54  &  1.286  &  572  &  3.48 \\
2.0  &  10  &  100.00  &  1.204  &  371  &  3.53 \\
3.0  &  1  &  1.00  &  1.579  &  1266  &  3.47 \\
3.0  &  1  &  4.64  &  1.421  &  1011  &  3.57 \\
3.0  &  1  &  21.54  &  1.293  &  686  &  3.65 \\
3.0  &  1  &  100.00  &  1.211  &  447  &  3.70 \\
3.0  &  10  &  1.00  &  1.579  &  1266  &  3.47 \\
3.0  &  10  &  4.64  &  1.421  &  1011  &  3.57 \\
3.0  &  10  &  21.54  &  1.293  &  686  &  3.65  \\
3.0  &  10  &  100.00  &  1.210  &  443  &  3.71\\
5.0  &  1  &  1.00  &  1.578  &  1484  &  3.70 \\
5.0  &  1  &  4.64  &  1.447  &  1241  &  3.77 \\
5.0  &  1  &  21.54  &  1.306  &  881  &  3.86 \\
5.0  &  1  &  100.00  &  1.212  &  572  &  3.93 \\
5.0  &  10  &  1.00  &  1.578  &  1484  &  3.70 \\
5.0  &  10  &  4.64  &  1.448  &  1241  &  3.77 \\
5.0  &  10  &  21.54  &  1.307  &  882  &  3.86 \\
5.0  &  10  &  100.00  &  1.211  &  568  &  3.93 \\
    \end{tabular}
    \caption{Subset of our evolutionary grid for which we compute spectra. These were chosen to explore the balance between flux from reflected-light and residual heat of formation.}
    \label{hyp_tab}
\end{table}

These ages and separations were chosen to overlap with known self-luminous planets, while also searching for the regime where both reflected light and residual heat of formation will contribute a significant amount of flux to the planet's optical spectrum. Since we sought the threshold where a 1.0 M$_J$ planet would be observable with WFIRST-CGI, we found ourselves considering ages right on the boundary between protoplanet and planet. A one to five Myr object could still be undergoing active accretion, and be surrounded by a circumplanetary disk, though the exact timescales should depend on the details of the circumstellar disk. There are several additional complications found in observing such young systems: accreting young stars are active into the optical until as late as $\sim$10 Myr and reflected light from the remnant transition disk can confuse planet searches. Bearing these caveats in mind, we suspect that the general trends and patterns pointed out in this work are still relevant, even if some of the masses and ages ought to include a circumplanetary disk component in their spectra, or consider the planet-circumstellar disk flux ratio in addition to the planet-star flux ratio. 

\subsection{Including Condensates}
NIR observations indicate that some of the self-luminous planets are likely to have dusty atmospheres (e.g. \citealt{madhu2011}; \citealt{Delorme2017}; \citealt{gravity2019}), so we also compute model spectra for HR 8799e with condensates. \cite{madhu2011} identified two best fit models for HR 8799d from their extensive grid-based search: (1) physically thick forsterite clouds and a 60-$\mu$m modal particle size and (2) clouds made of 1-$\mu$m pure iron droplets and 1\% supersaturation. We chose to use the best-fit cloud model with forsterite particles because silicates are generally considered to be responsible for the L-T transition in brown dwarfs. However, this particular cloud formulation should be considered one of many possible options consistent with current observations, not the most likely scenario. A variety of cloud models incorporating a wide range of geometrical and optical thicknesses have been applied to this system and other substellar atmospheres (see \citealt{Ackerman2001}; \citealt{Marley2002}, \citealt{Tsuji2005}, \citealt{Burrows2006}, \citealt{Helling2008}, and \citealt{burrows2011} for some examples of brown dwarf dust and cloud treatments). The conclusion has been that clouds almost certainly exist at high altitudes in these atmospheres, but the cloud composition, cloud particle sizes and gas-phase metallicity are not well-constrained by the available NIR data. Optical observations, like those modeled in this work, combined with NIR spectra, will hopefully provide the best possible constraints on the condensates in self luminous planets and brown dwarf companions. 

\begin{figure}
    \centering
    \includegraphics[width=\textwidth]{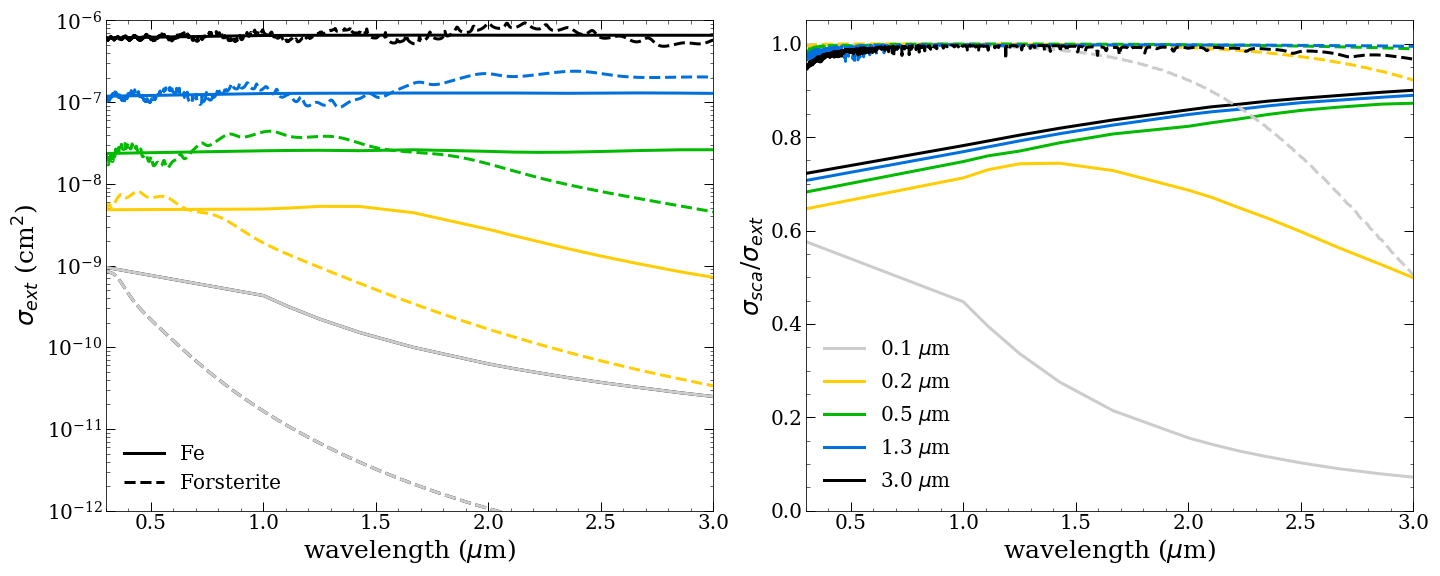}
    \caption{Optical properties of iron and forsterite, two candidate species for the dusty condensates inferred to be present in HR 8977c, d, and e. The extinction cross section (\textbf{left}) and single-scattering albedo (\textbf{right}) are shown as a function of wavelength for several particle sizes. These are calculated using Mie theory, so assuming homogeneous spherical particles. Complex indices of refraction for the species were taken from \citealt{Kitzmann2018}. In practice, the atmospheres may contain aggregate particles made up of both condensates which will exhibit an intermediate behavior between the pure iron or pure forsterite case.}
    \label{fig:opt_props}
\end{figure}

\begin{figure}
    \includegraphics[width=0.5\textwidth]{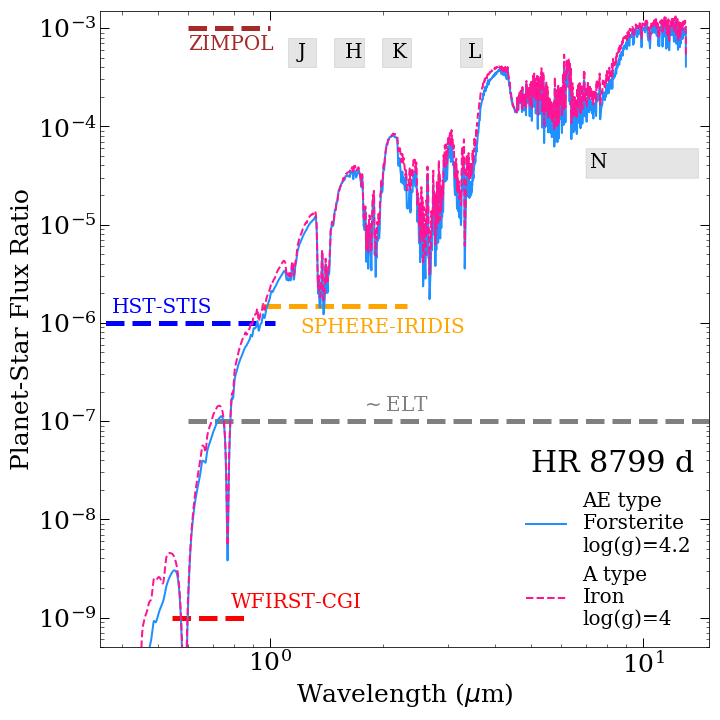}
    \caption{The spectra for the two best-fit models to HR 8799 d observations from \cite{madhu2011}, overlaid with the contrast capabilities and wavelength coverage of some current and future direct-imaging instruments. The models from \cite{madhu2011} were converted to a planet-star flux ratio using the spectra for an A5V star from the Pickles Stellar Atlas \citep{Pickles1998} extended to longer wavelengths with a black body of the appropriate temperature.}
    \label{fig:madhu}
\end{figure}

Figure \ref{fig:opt_props} compares the optical properties of iron and forsterite as a function of wavelength and particle size. Looking at the extinction coefficients in the left panel of Fig. \ref{fig:opt_props}, one can see how the two condensate species can be made to look like each other in a narrow wavelength range by varying particle size or the number of particles. Figure \ref{fig:madhu} shows spectra for the two best fit models of \cite{madhu2011}. This demonstrates how different condensate species, particle sizes and planet properties can be degenerate when observations cover only a narrow wavelength range (in this case the H and K bands), but become distinguishable when more wavelengths are considered in concert. If planets are close to their host star and cool enough, reflected light can contribute significant flux at wavelengths $<$ 0.6 $\mu m$. Looking at the single-scattering albedos shown in the right panel of Fig. \ref{fig:opt_props}, one can see that silicates are much more reflective than iron for most particle sizes and wavelengths. In a case where both reflected light and NIR observations are available, these two candidate species would be even less degenerate.

It is not yet known what condensate species are present in exoplanet atmospheres. For this reason, when we include condensates in the atmospheres of our hypothetical planets, we remain agnostic to the precise species and instead parameterize their effects with the addition of a purely scattering gray opacity (in cm$^2$/g) to the other opacity sources included in our model. We add the extra scattering opacity uniformly throughout the whole atmosphere. The species and altitude of the cloud condensates will play a critical role in determining how bright a planet is in the optical, and in determining the albedo at all wavelengths (\citealt{Marley1999}; \citealt{sudarsky2000}). This can be understood by considering the relative contributions of Rayleigh scattering, cloud scattering, gaseous absorption, and any absorption from the clouds to the total atmospheric opacity. If an atmosphere is completely cloud-free, then, generally, Rayleigh scattering will be the dominant scattering opacity source at wavelengths shorter than 0.6 $\mu$m, leading to a higher geometric albedo in the optical, while gaseous absorption will be dominant at longer wavelengths, leading to a lower geometric albedo in the NIR. A high cloud can generally raise a planet's geometric albedo at all wavelengths, since light will scatter back outwards without encountering as much absorbing gas as a low cloud or no cloud at all. However, it is also possible to form condensates which darken the optical appearance of the planet relative to purely Rayleigh scattering atmospheres. If a cloud species contributes even a small amount of absorption at optical wavelengths (e.g. Iron or Na$_2$S, \citealt{Morley2012}), it can darken an atmosphere as multiply-scattered light has a higher probability of being absorbed rather than escaping \citep{mayorga2019}. If a cloud species forms very low in the atmosphere, then it will have little or no effect on the geometric albedo. 

Thus, our choice of a purely scattering gray opacity encapsulates the behavior of only a subset of those cloud species and geometric extents which are physically plausible. Depending on the temperature of an atmosphere, different species may condense at different altitudes under equilibrium conditions. For planets in the range of 900-3000 K, condensates will be silicates like forsterite, soots, or iron. For planets on the hotter end ($>$1500 K), condensates will form high up in the atmosphere. For planets on the cooler end of that range, condensates will form deeper in the atmosphere. For slightly cooler planets (650-800 K), condensates might include compounds like Na$_2$S, KCl and ZnS \citep{Morley2012}, or nothing at all. Planets around 350-650 K will likely be clear, with only Rayleigh scattering. Water will condense for cool planets ($<$350 K), and, for the coolest mature planets ($<$150 K) ammonia will condense \citep{sudarsky2000}. Photochemical hazes may also form and play a significant role for some planets. As young planets cool, they will go through periods of time when they exhibit different geometric albedos. A detailed self-consistent study of the effect of condensates and planet-star separation throughout the evolution of young giant planets is certainly warranted, but beyond the scope of this work.
 
\section{Results: Optical Spectra of Known Systems Observable with WFIRST-CGI} \label{sec:known}

\begin{figure}[htb!]
\includegraphics[width=\textwidth]{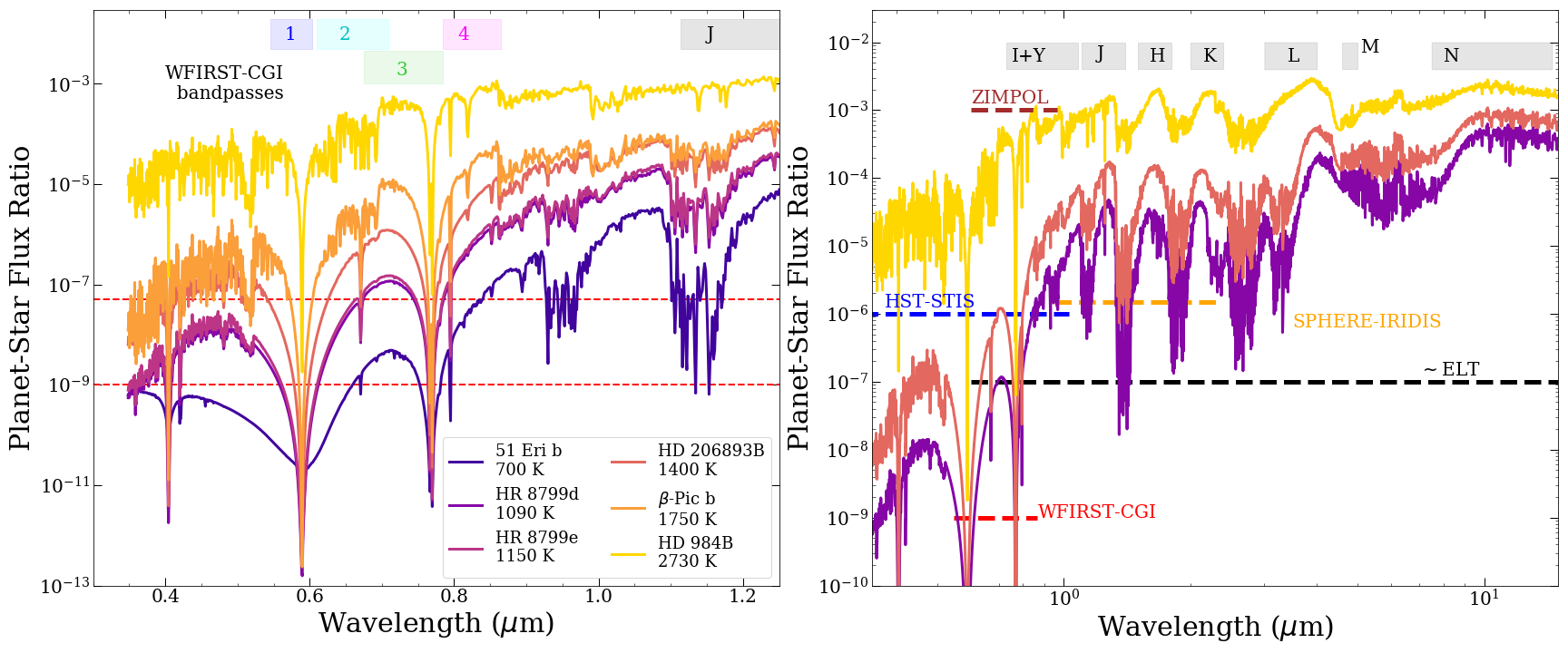}
\caption{\label{known_spectra}The panel on the \textbf{left} shows optical spectra for six known systems with clear atmospheres and solar abundances. We highlight the wavelength coverage of WFIRST-CGI bands 1, 2, 3, and 4 with colored rectangles. Part of the NIR J band is also shown in gray for reference. Red dashed lines indicate the mission requirement contrast of WFIRST-CGI at 5$\times$10$^{-8}$ and the current best estimate contrast at 10$^{-9}$. The \textbf{right} panel shows a selection of three of the spectra from the left panel, extended to longer wavelength ranges and overlaid with the contrast and wavelength coverage of some existing and planned direct-imaging instruments. The atmospheric windows are denoted with gray rectangles across the top of the figure.}
\end{figure}

The top left panel of Fig. \ref{known_spectra} shows model spectra for the five self-luminous companions with working angles and host star magnitudes most suitable for observations with WFIRST-CGI at optical wavelengths. The first simple, but significant, result is that objects with temperatures of $\sim$1000 K or above (e.g. HD 984B, $\beta$-Pic b, HD 206893B, and HR 8799e), are within the observing capabilities of WFIRST-CGI, even in its descoped form (required and predicted contrast limits for WFIRST-CGI are denoted by the red dashed lines). 51 Eri b, will also be observable if WFIRST-CGI reaches its predicted performance of a 10$^{-9}$ contrast floor. Optical observations of self-luminous substellar companions have long been ignored because they are so much dimmer than the NIR, but this new wavelength range will become very accessible with future instruments. It is also worth noting that some of these objects may be detectable with an existing instrument, HST-STIS. The top right panel of Fig. \ref{known_spectra} shows a selection of the objects compared to the contrast limits and wavelength coverage of SPHERE-ZIMPOL, SPHERE-IRIDIS, HST-STIS, an approximate ELT with second generation optical AO, and WFIRST-CGI. Companions as hot as HD 984B (2730 K) would be observable with the full HST-STIS wavelength coverage. Planets around the temperature of $\beta$-Pic b (1750 K) would be observable with HST-STIS as blue as $\sim$0.65 $\mu$m. Planets around 1000 K could be observed as blue as $\sim$0.8 $\mu$m. With these temperatures in mind, and noting that the inner working angle of HST-STIS is around 0.5$''$, it seems feasible that HIP 65426 b, HD 1160B, HR 3549B, HD 95086b,  $\kappa$-And b, HR 8799 d, and HR 2562B could be observed with some portion of the HST-STIS wavelength range. More detailed models for those systems and HST-STIS specific exposure time calculations would be needed to say for sure.

\begin{figure}[htb!]
\includegraphics[width=0.66\textwidth]{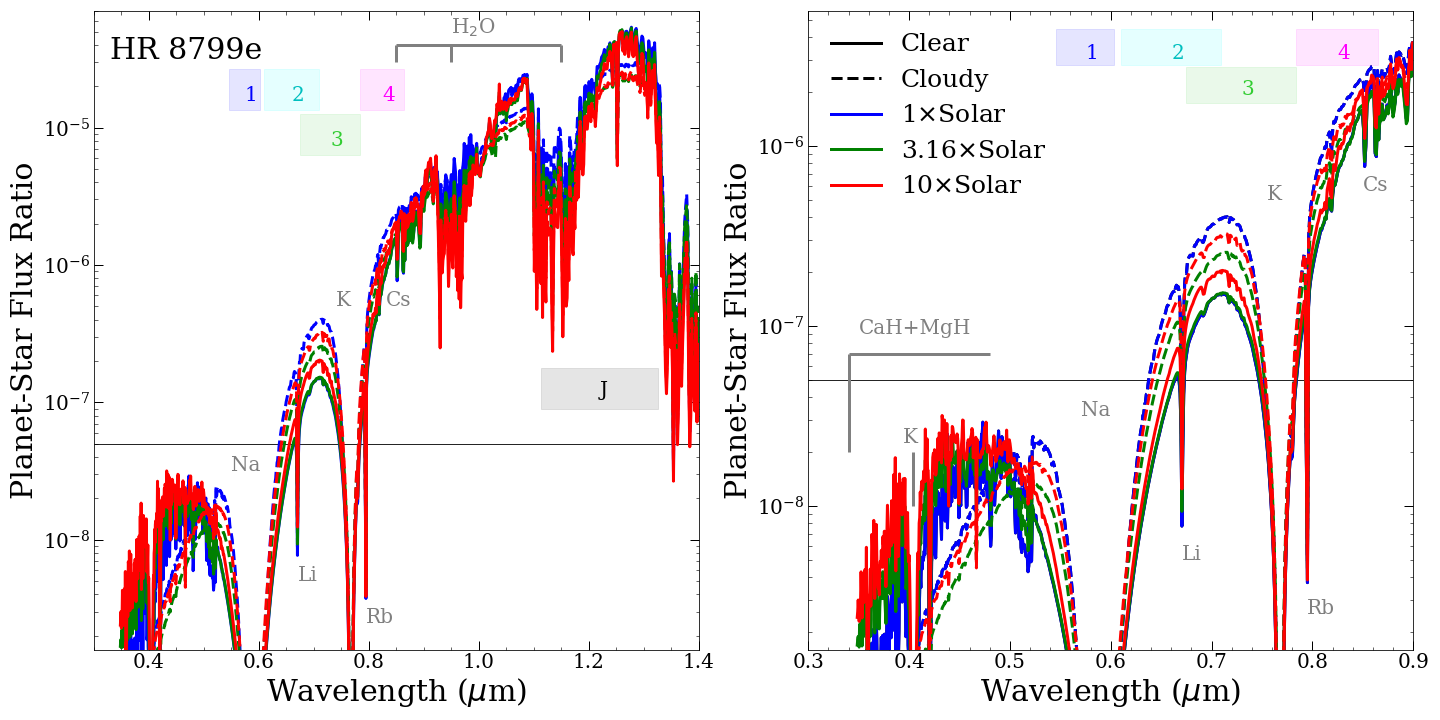}
\includegraphics[width=0.33\textwidth]{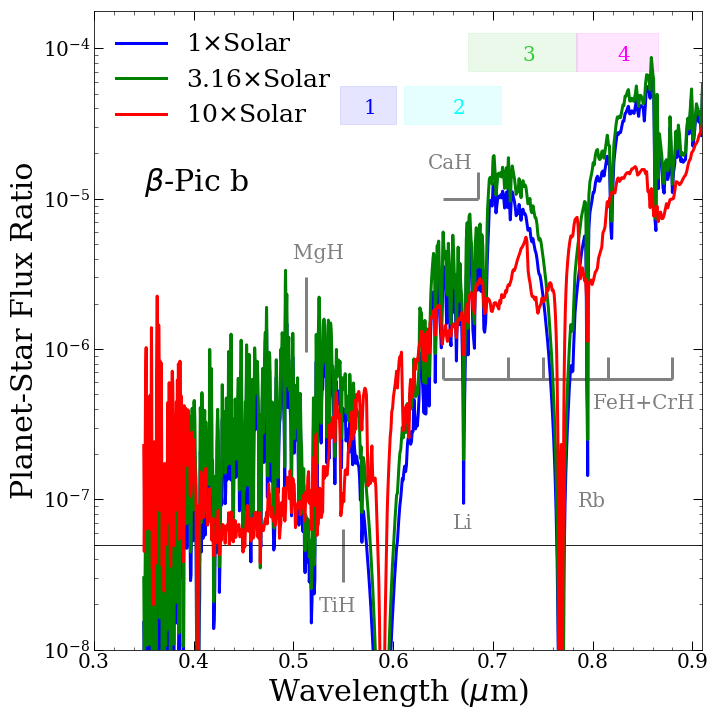}
\caption{\label{known_spectra_metallicity}The panels on the \textbf{left} and in the \textbf{center} show spectra for HR 8799e comparing clear (solid lines) and cloudy (dashed lines) atmospheres at solar, 3.16$\times$solar and 10$\times$ metallicities, with NIR wavelengths included on the left and optical wavelengths only in the center. The \textbf{right} panel shows spectra for $\beta$-Pic b at solar, 3.16$\times$solar and 10$\times$solar metallicities, highlighting how the shape of the 0.77-$\mu$m potassium doublet will change with metallicity, and significantly increased optical opacity at 10$\times$solar. In all panels we use colored rectangles to note the wavelength coverage of WFIRST-CGI and black lines indicate the mission requirement contrast of WFIRST-CGI at 5$\times$10$^{-8}$.}
\end{figure} 

These optical-wavelength-range spectra contain some interesting features, similar to those seen in L and T dwarfs. A potassium resonance doublet will dominate the WFIRST-CGI bandpasses at 0.77 $\mu$m, covered towards the blue side by spectral bandpass 3 and towards the red side by imaging bandpass 4. There is also a strong sodium resonance doublet at 0.59 $\mu$m, which falls mainly within the shortest wavelength imaging bandpass, with some of the red wing of the feature reaching into spectral bandpass 2.  As atmospheres cool below $\sim$750 K (e.g. 51 Eri b), Rayleigh scattering contributes significant amounts of reflected light blueward of 0.6 $\mu$m. Other deep narrow features within the 0.5-0.9 $\mu$m WFIRST-CGI wavelength range include lithium, rubidium, and additional weaker sodium and potassium features. In atmospheres with temperatures above 1400 K (e.g., HD 206893B, $\beta$-Pic b, and HD 984B), metal hydrides cause the jagged appearance. In cooler atmospheres, once the hydrides rain out, the photosphere moves deeper in the atmosphere and we see greater collisional broadening of the sodium and potassium doublets. The resolution of the spectra shown here are R$\sim$740. This is much higher than the average WFIRST-CGI resolving power of R$\sim$50, so much of the fine structure seen in the spectra will be unresolved.

Comparing models with non-solar metallicities indicates that obtaining spectra in the optical wavelength range could help to constrain the metallicities of the hotter self-luminous planets, or cooler planets with silicate clouds. The right-most panel of Fig. \ref{known_spectra_metallicity} compares spectra of $\beta$-Pic b with solar, 3.16$\times$solar and 10$\times$solar metallicity focusing on just the optical wavelengths. As metallicity increases to 10$\times$solar, increasing metal hydride opacity siginificantly alters the shape of $\beta$-Pic b's spectrum in the WFIRST-CGI bands. The left panel and center panels of Fig. \ref{known_spectra_metallicity} show HR 8799e at solar, 3.16$\times$solar and 10$\times$solar metallicity, both with forsterite clouds (dashed lines) and without (solid lines). Considering first just the clear atmospheres for HR 8799e, one can see that the 1 and 3.16$\times$solar spectra hardly deviate at all from each other for wavelengths between 0.55 and 0.9 $\mu$m. At just 1150 K, HR 8799e is not hot enough for as many metal hydrides to be present. However, when forsterite clouds are present high in the atmosphere, a stronger dependence on metallicity is seen. Clouds move the photosphere to a higher altitude, narrowing the collisionally-broadened 0.59-$\mu$m sodium and 0.77-$\mu$m potassium features and causing a higher flux for wavelengths between 0.5 $\mu$m and 0.9 $\mu$m. With forsterite clouds included, the planet's brightness around 0.7 $\mu$m increases by a factor of 2-4, depending on metallicity. Looking at the left panel, one can see how clouds suppress regions of emission and lessen absorption in general. While not a comprehensive demonstration of all possible cloud formulations, these models indicate that the wavelength coverage of WFIRST-CGI will be strongly affected by the presence or absence of condensates in young giant planet's atmospheres. 

\begin{figure}
    \centering
    \includegraphics[width=\textwidth]{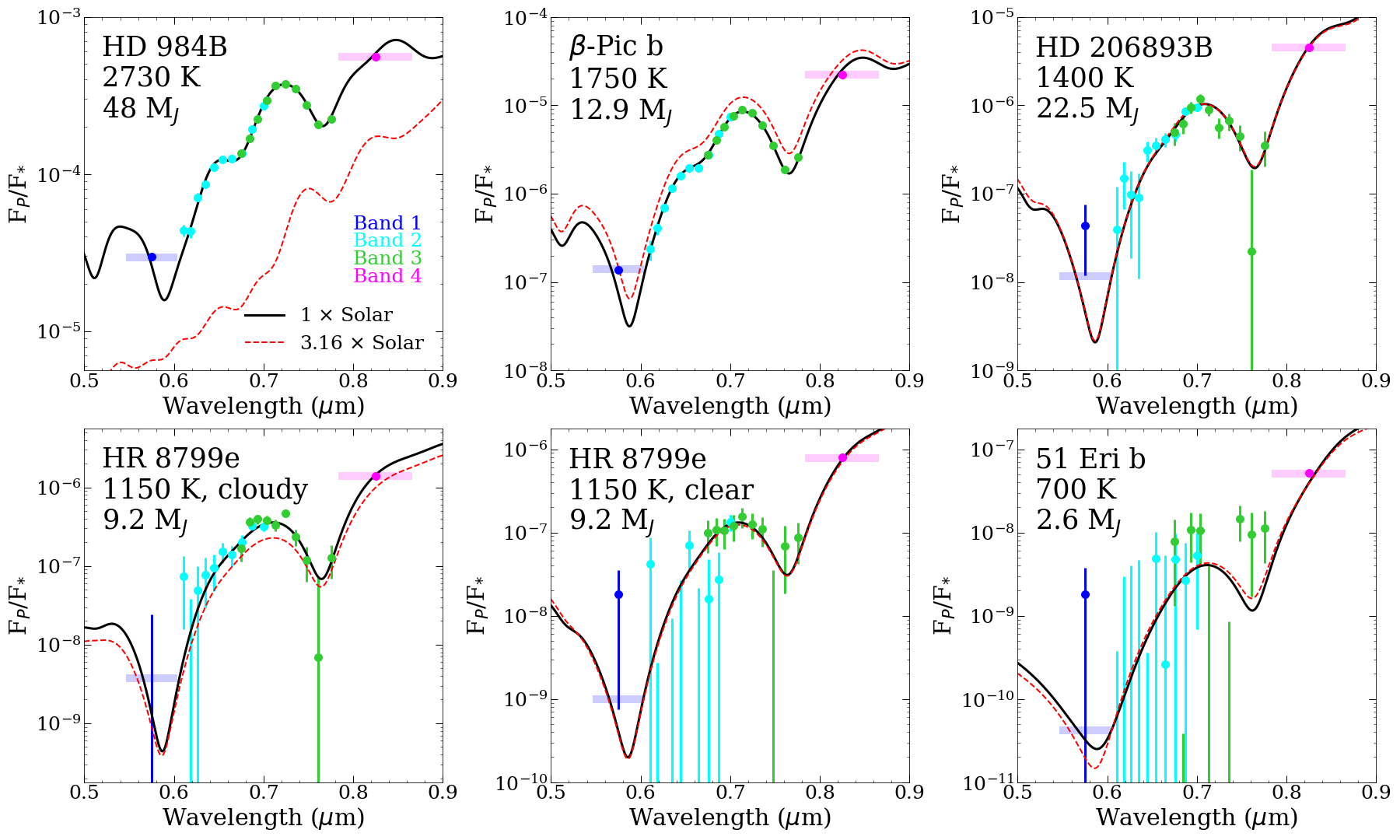}
    \caption{\label{fig:simobs}Simulated WFIRST-CGI observations of HD 984B, $\beta$-Pic b, HD 206893B, HR 8799e with a cloudy atmosphere, HR 8799e with a clear atmosphere, and 51 Eri b. Exposure times vary, but in all cases we limited them to below the maximum exposure times of 500 hours for spectra and 100 hours for imaging. We assume a noise from one exozodi's worth of background dust scattering light in all cases. In reality, $\beta$-Pic has an extremely bright debris disk, which will most likely drown out the signal modeled here. Accounting for the extremely bright disk around $\beta$-Pic, these simulated observations show that HD 984B, HD 206893B, and HR8799e are the best targets for WFIRST-CGI to obtain spectra.}
\end{figure}

Figure \ref{fig:simobs} shows simulated WFIRST-CGI observations of the five planets which will be observable in the $\lambda$=0.575-0.785 $\mu$m bands. Exposure times and signal-to-noise ratios are calculated using the methods outlined in \cite{Lacy2019}, modified to account for the varying spectral resolution of the slit-prism spectrograph that has replaced the integral-field spectrograph, and assuming the coronagraph contrast and other instrument specifications are set at mission requirements \citep{actualMissionReqs}. In all these systems we modeled the noise contribution from exozodiacal dust as equivalent to the amount of dust in our own solar system. The hottest planet, $\beta$-Pic b, and even hotter brown dwarf companion, HD 984B, need only sub-hour-long exposures to attain high signal-to-noise ratio measurements. HD 206893B also needs only a moderate exposure time. HR 8799e needs exposure times more comparable to the upper limits of what is reasonable for optimal dark-hole digging. 51 Eri b cannot attain reasonable signal-to-noise ratios even with the maximum feasible exposure time. Our assumption of a disk background equivalent to one exozodi is certainly not an accurate representation of the extremely bright debris disk around $\beta$-Pic, with a planet-star flux ratio of around 10$^{-3}$ \citep{Apai2015}. In this system, it is possible that reflected light from the disk (M$_V$ $\sim$16.7 at the location of $\beta$-Pic in 2015) will completely overwhelm the signal from $\beta$-Pic b in WFIRST-CGI's wavelengths. Detecting and characterizing exozodiacal dust and debris disks around potential targets for future exoearth searches is another likely application for WFIRST-CGI's extended mission phase \citep{2019summary}.

For planets with effective temperatures above 1700 K and clear atmospheres, the signal-to-noise ratio looks adequate to distinguish between solar and 10$\times$solar metallicity, assuming metal hydrides, VO and TiO are present in equilibrium abundances. At cooler temperatures or lower metallicities, sensitivity to metallicity is seen in the photometric bandpass 4, and the shape of the potassium absorption feature covered by spectral bandpass 3, but it is not clear that these won't be degenerate with other planet properties given the low resolution of the observations. Based on the example of HR 8799e, it appears that condensates, which are expected to be present in the atmospheres of HR 8799e and HD 206893B, may leave imprints large enough to be measured by WFIRST-CGI. A more detailed study folding in uncertainties in temperature, surface gravity, and the effects of possible condensates needs to be done before one can state the precision with which WFIRST-CGI can measure metallicities or constrain cloud properties for young giant planets. Again, we emphasize that best constraining power will come from combining optical wavelength observations with available NIR measurements.

\section{Results: Hypothetical Systems}\label{sec:hypothetical}
In this section we shift from the earlier focus on WFIRST-CGI targets and imminent mission planning, to an exploration of the possible insights to be gained from observing young giant planets with possible upcoming missions aiming to search for exoearths. We focus especially on determining the combinations of age, mass, and planet-star separation where both reflected light and residual heat of formation will contribute significant optical flux to the observed planet spectra.

\subsection{Irradiated Evolutionary Model Grid and Spectra}

\begin{figure}[htb!]
\includegraphics[width=\textwidth]{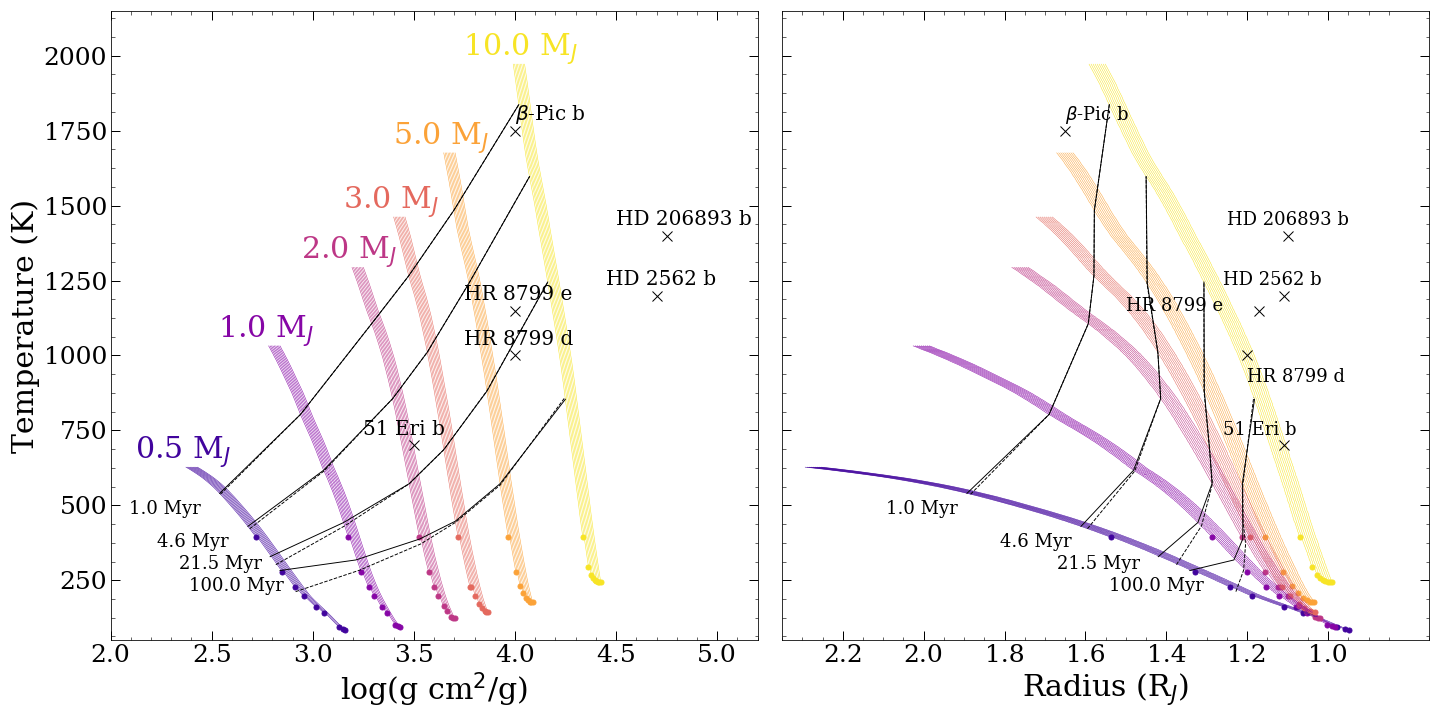}
\caption{\label{ev_tracks}Evolutionary tracks for a selection of masses in the plane of temperature and surface gravity (left), and in the plane of temperature and radius (right). All object masses were computed for a variety of planet-star separations (0.5, 1, 1.5, 2, 3, 4, 10, 15, and 20 AU) assuming stellar irradiation from a sun-like star. The different separations are shown with an artificial offset in the x direction, because the tracks lie essentially on top of each other for the early stages of their evolution, only deviating when irradiation begins to compete with residual heat of formation. Dots indicate the terminal temperature and surface gravity or temperature and radius, showing that, regardless of mass, objects eventually cool to a temperature set by their separation from the host star, while the final radii will vary with both separation and mass. Known systems modeled in \S \ref{sec:known} WFIRST-CGI are marked and labeled. We also include isochrones spanning 1 to 100 Myr. The solid black lines correspond to separations of 1 AU and the dashed black lines correspond to separations of 10 AU.}
\end{figure}

First, we computed evolutionary models for a grid of planet masses and planet-star separations. These models provide effective temperatures, radii and surface gravities that can be used as input for our spectral code. The resulting tracks in temperature-surface gravity and temperature-radius space are shown in Fig. \ref{ev_tracks} for the full grid of models. Early on there is not a strong dependence on planet-star separation. Objects with different masses at the same separation converge to the same temperature eventually, but to different radii and surface gravities. In Fig. \ref{ev_tracks}, this is evident as the terminal dots line up horizontally for planets at 0.5, 1.0, and 1.5 AU separations. At larger separations, the larger mass objects have not had sufficient time to cool to their equilibrium temperatures within the timeframe of our models. Closer-in planets at a given mass have larger radii, and thus smaller surface gravities (note the reversed x-scale in the right hand side). There is some overlap in parameter space between our evolutionary grid and the known systems modeled in the previous section, but we extend our grid to smaller, older planets in addition to the known ones, because later missions will make these observable (as demonstrated in Fig. \ref{fig:tongue}). Isochrones are shown to guide the eye towards the subset of this grid for which we compute spectra: 1.0, 2.0, 3.0, and 5.0 M$_J$, at 1.0 and 10.0 AU, and at 100.0, 21.5, 4.64, and 1.0 Myr.  

\begin{figure}[htb!]
\begin{center}
\includegraphics[width=0.8\textwidth]{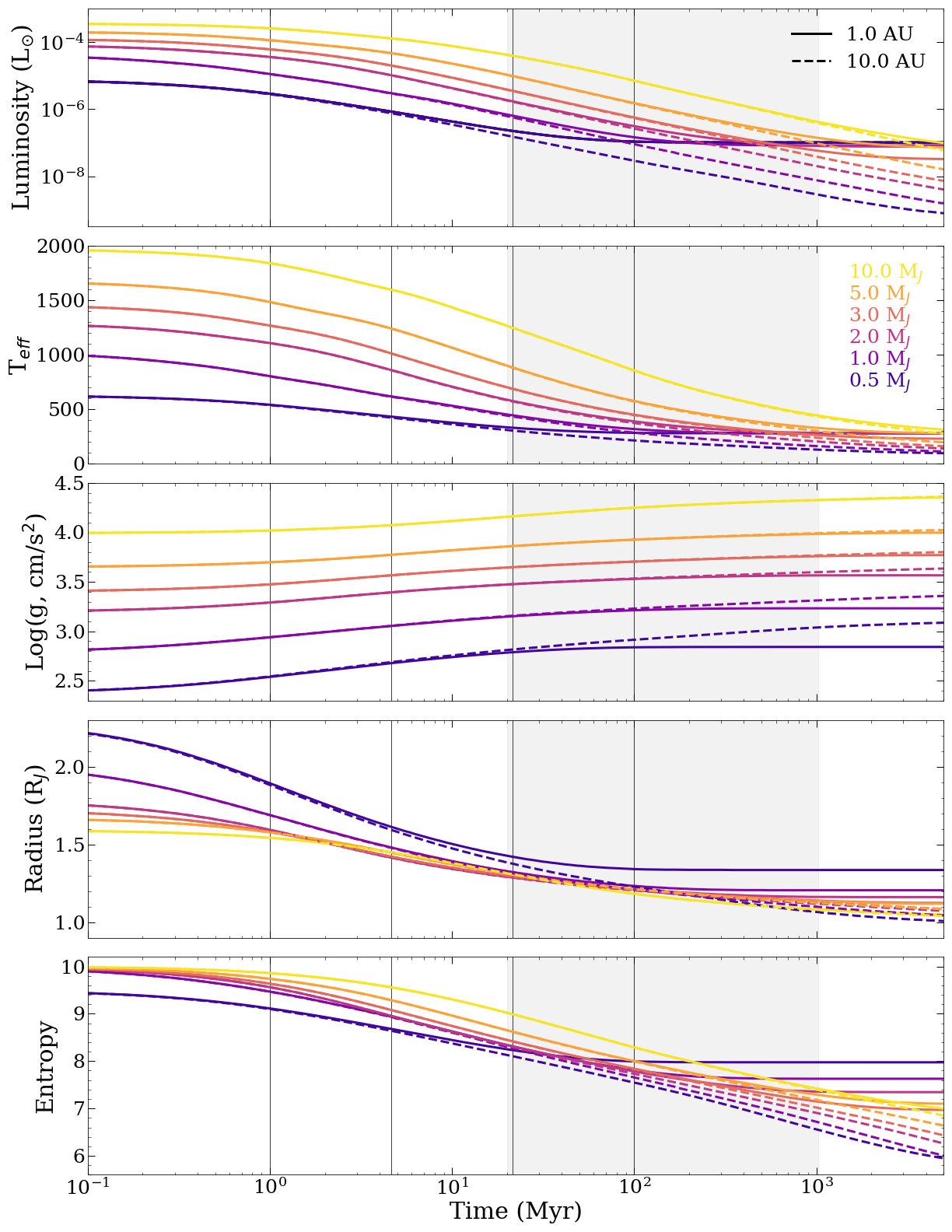}
\end{center}
\caption{\label{ev_v_time}Evolution with time for a variety of planet masses (denoted by colors) and separations (denoted by line style). Smaller planets experience a larger effect from the variation in planet-star separation within the time frame modeled here, but the planet-star separation determines the final effective temperature, regardless of the mass of the planet, given enough time. The shaded gray region marks the ages spanned by the known directly-imaged planets modeled in the preceding section. The vertical black lines note 1, 4.65, 21.5 and 100 Myrs.}
\end{figure}

Figure \ref{ev_v_time} shows the time evolution of bolometric luminosity, effective temperature, surface gravity, radius, and specific entropy per baryon per $k_{B}$ for the different masses at two separations.   Again, it is apparent that early in their evolution planets are not strongly affected by incoming irradiation. It is only as planets approach the equilibrium temperature for their separation that the tracks for 1 and 10 AU deviate (compare dashed vs dotted lines of the same color). Closer-in planets end with a higher temperature, and so correspondingly higher luminosity, higher entropy, larger radius and smaller surface gravity. Smaller planets cool faster, so they exhibit a dependence on planet-star separation at a younger age than more massive planets. 

\begin{figure}[htb!]
\begin{center}
\includegraphics[width=0.85\textwidth]{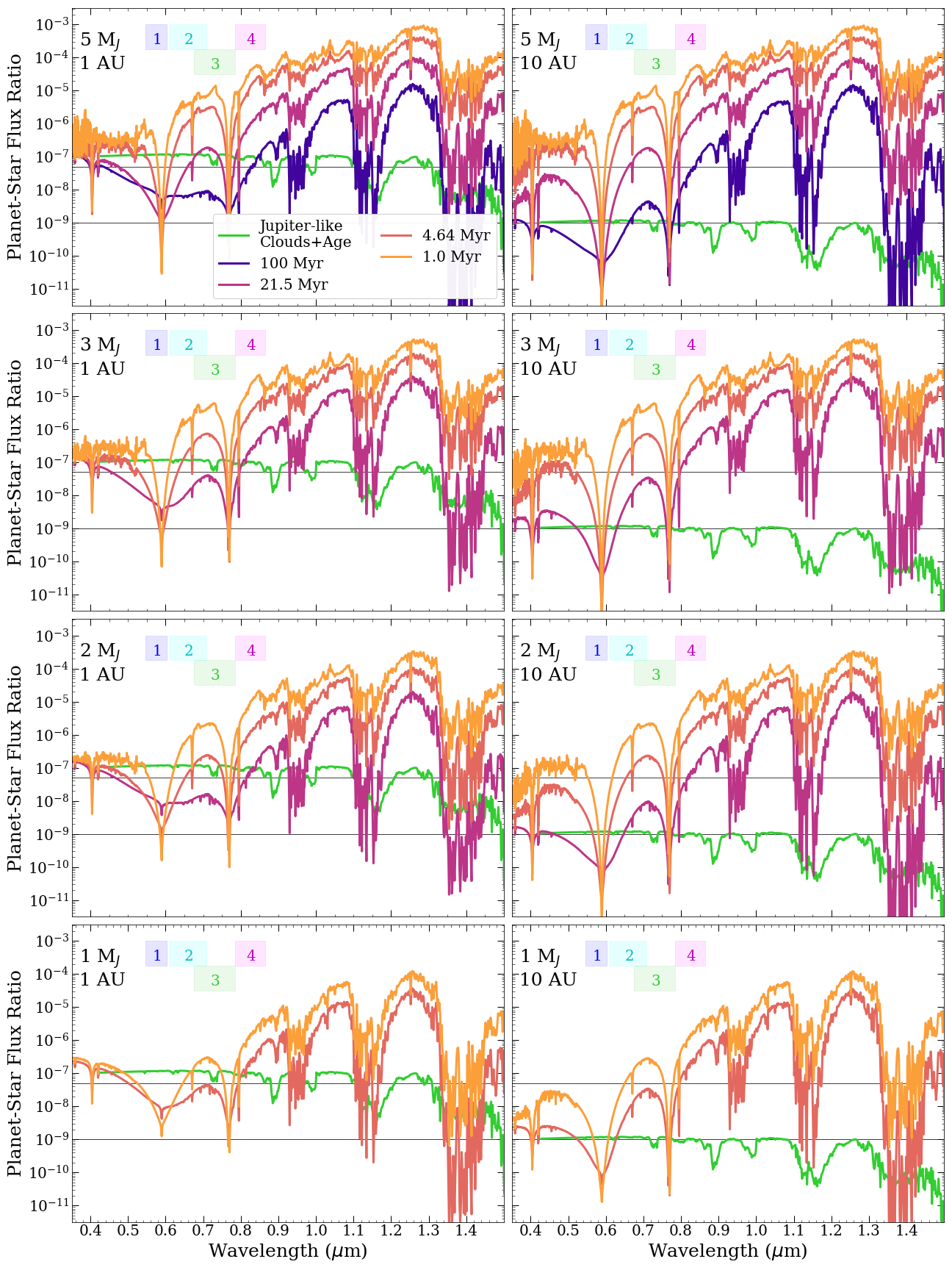}
\end{center}
\caption{\label{hypothetical_planets}Spectra for selected stages in our evolutionary model grid. Each panel represents a different mass and separation, with spectra at up to four ages (100.0 Myr, 21.5 Myr, 4.64 Myr, and 1.0 Myr). The left column has separations of 1 AU, while the right column has a separation of 10 AU. Each row corresponds to a different mass, from the top: 5, 3, 2, and 1 M$_{J}$. The bandpasses of WFIRST-CGI are denoted by colored rectangles and horizontal black lines show the WFIRST-CGI required contrast ratio of 5$\times$10$^{-8}$ and the engineer's target performance 10$^{-9}$. We also include a Jupiter-like reflected light spectrum in green for comparison. Table \ref{hyp_tab} lists the corresponding temperatures, radii, and surface gravities used in computing these spectra.}
\end{figure}

Spectra for a representative sample of masses, ages, and separations are shown in Fig. \ref{hypothetical_planets}. Each row corresponds to a different planet mass and each column a different planet-star separation. Purple, magenta, coral, and orange lines show spectra for 100, 21.5, 4.64, and 1 Myr, respectively. The temperatures, radii, and surface gravities used for each spectrum can be found in Table \ref{hyp_tab} in \S \ref{sec:models}. For comparison, we added a Jupiter-like spectrum scaled as if orbiting at 1 and 10 AU respectively, shown in green. We now highlight some patterns and results, illustrated by this selection of spectra:

\begin{itemize}
    \item Young giant planets can be brighter than a mature Jupiter-like planets in between the sodium doublet at 0.59 $\mu$m and the potassium doublet at 0.77 $\mu$m, and at wavelengths shorter than $\sim$0.5 $\mu$m. 
    \item At 1 AU, a cool cloudy Jupiter will outshine a clear young planet once it cools below $\sim$750 K. Out at 10 AU, young giant planets remain brighter than Jupiter-like planets in the optical for all the spectra we computed (down to 575 K). 
    \item Planets with clear atmospheres warmer than $\sim$700 K will be detectable in the optical for a mission with contrast limits on the order of 10$^{-7}$. For a one-Jupiter-mass planet, this means an age of around 5 Myr or younger. Most known self-luminous planets and substellar companions fall under this category (with the possible exception of 51 Eri b).
    \item Planets with clear atmospheres warmer than $\sim$500 K will be detectable in the optical for a mission with contrast limits on the order of 10$^{-9}$. For a one-Jupiter-mass planet, this means an age of around 10 Myr or younger. Self-luminous planets a bit older or less massive than currently known directly-imaged planets will be bright enough to be observed in the optical with WFIRST-CGI if it achieves the goal contrast ratio of 10$^{-9}$. 
    \item The age at which reflected light contributes to a planet's spectrum spectrum depends on the planet-star separation and planet mass. Closer-in planets can reflect more light because they receive more stellar irradiation to begin with. Smaller mass and closer-in objects have larger radii at a given age, providing more surface area to reflect incoming stellar irradiation. Larger mass objects cool more slowly, and have less of a radius dependence on planet-star separation.
    \item At a 1 AU planet-star separation, reflected light is seen blueward of 0.4 $\mu m$ for all these masses and ages (1-5 Jupiter-masses, 1-100 Myr). The reflected light comes from H$_2$ and He Rayleigh scattering, since we assume the atmospheres to be clear. 
    \item At 1 AU, as planets cool below $\sim$750 K, reflected light begins to dominate the shape of their spectra blueward of $\sim$0.7 $\mu m$.
    \item At 10 AU, clear planet atmospheres need to cool down to $\sim$600 K for reflected light to contribute significantly blueward of $\sim$0.65 $\mu m$.
\end{itemize}

\begin{figure}[htb!]
\begin{center}
\includegraphics[width=0.85\textwidth]{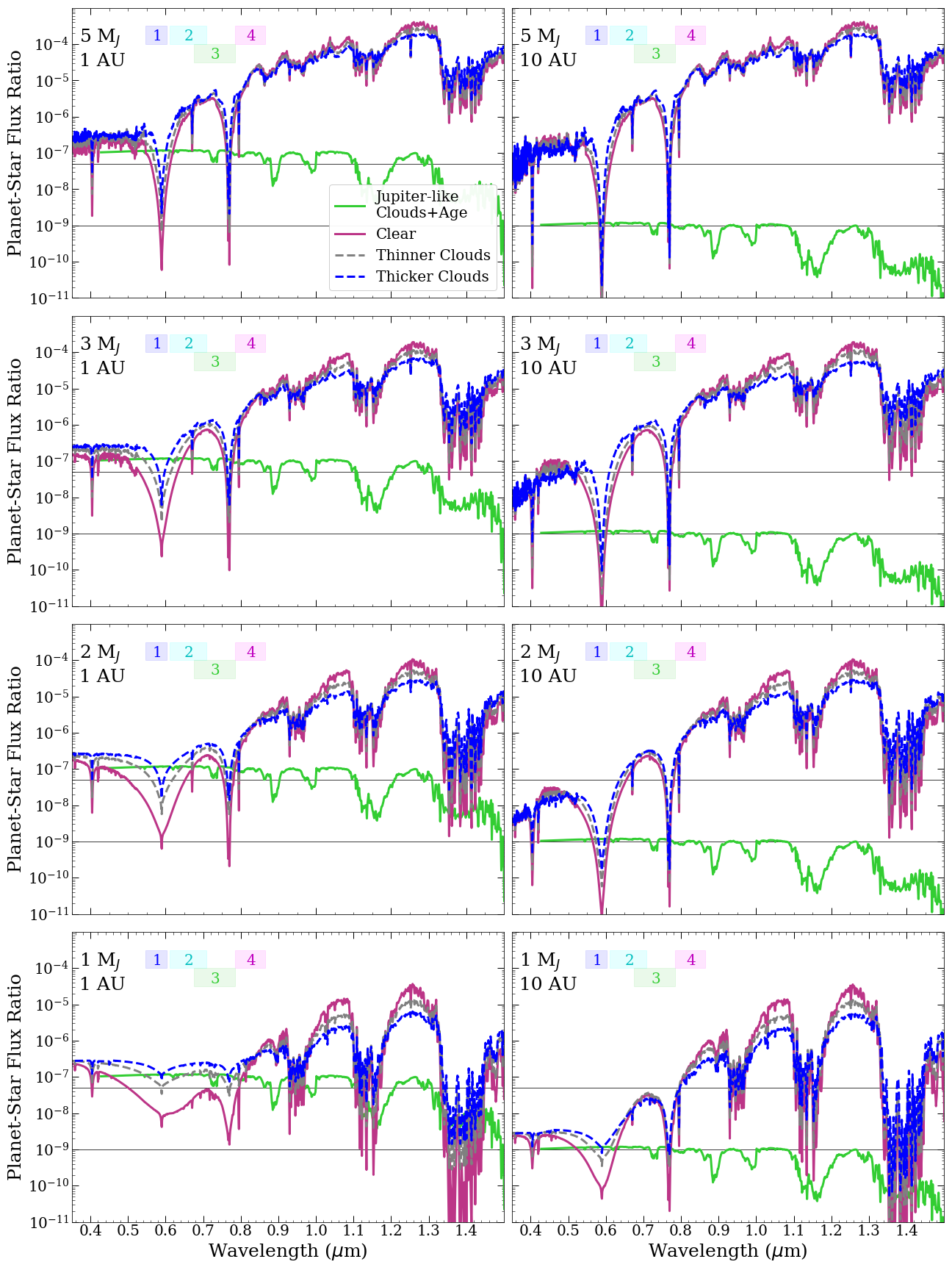}
\end{center}
\caption{\label{various_condensates}4.64-Myr spectra for the same masses and separations shown in Fig. \ref{hypothetical_planets}, now comparing clear atmospheres (magenta line) to atmospheres with an additional 100\%-scatterering opacity. The gray dashed line corresponds to a scatterer with $\kappa_s$=0.002 cm$^2$/g and the dashed blue line corresponds to a scatterer with $\kappa_s$=0.01 cm$^2$/g. Again, the green line shows a Jupiter-like planet at the same separation, colored rectangles denote the WFIRST-CGI bandpasses, and black horizontal lines mark the WFIRST-CGI baseline requirement and target contrast levels. Including additional scatterers shifts the wavelength where reflected light contributes significantly towards redder wavelengths.}
\end{figure}

As discussed previously it is likely that condensates will be present in some young giant planet atmospheres. Figure \ref{various_condensates} shows all of the 4.64 Myr objects from Fig. \ref{hypothetical_planets}, now with an extra purely scattering gray opacity added. We compare a clear atmosphere in maroon with a ``thicker" cloud of opacity 0.01 cm$^{2}$/g in blue and ``thinner" cloud opacity of 0.002 cm$^{2}$/g in gray. As we have already seen for HR 8799e, condensates diminish the Na and K absorption raising the planet-star flux ratio shortward of 0.8 $\mu$m for planets at all separations. For the planets at 1 AU, the additional scattering opacity raises the amount of reflected light by a factor of $\sim$2-10 relative to Rayleigh alone. At longer wavelengths the extra scatterer diminishes the emission peaks and lessens the depths of CH$_4$ and H$_2$O absorption bands. We emphasize again, that depending on the optical properties of any condensate species, their presence could either raise the planet's reflection as seen here, or absorb more light, darkening it. 

\subsection{Reflected Light for Self-Luminous Planets}
\label{sec:disentangle}

Planet-star flux ratios in reflected light are commonly described by the following relation:
\begin{equation}
    \left(\frac{F_{P}}{F_{*}}\right)_{ref} = \Phi(\lambda,\alpha)A_{g}(\lambda)\left(\frac{R_{P}}{r}\right)^2 \rm,
\end{equation} 
where $\Phi$ is the phase function, $\lambda$ is the wavelength of light, $\alpha$ is the phase angle describing the relative locations of the star, planet, and observer, $A_{g}$ is the geometric albedo, $R_P$ is the radius of the planet, and $r$ is the planet-star separation corresponding to $\alpha$. The phase function describes how much reflected light will scatter towards the observer given the phase angle. It is usually normalized such that it is equal to one when the planet is fully illuminated. The albedo and scattering properties of a planet's atmosphere reveal information about particle compositions, particle sizes, and depths of any scattering condensate layers (\citealt{Marley1999}; \citealt{sudarsky2000}; \citealt{cahoy2010}; \citealt{Seager2010}). Actual phase functions for solar system bodies have been found to differ from both a Lambertian phase function (the analytic solution for an isotropically scattering sphere) and from the phase function for a homogeneous Rayleigh-scattering atmosphere (\citealt{madhu2012}; \citealt{mayorga2016}). Measuring exoplanet's phase functions can thus inform us about the condensates present in their atmospheres, and is also important for planning and interpreting direct imaging observations of reflected light from mature planets. 

For the youngest hottest planets, any reflected light will be entirely overpowered by flux from the residual heat of formation. However, as planets age, they contract and cool. The relative contributions of residual heat of formation and reflected light to their optical spectra will change over time accordingly. Understanding the balance between these two sources of flux can help us to interpret the selection of spectra shown in Fig. \ref{hypothetical_planets} and to identify under what conditions we expect spectra to contain significant contributions from both reflected light and thermal emission. These objects present an opportunity to measure radii, geometric albedos and phase functions without the degeneracies that plague similar measurements for mature planets. Reflected light observations alone contain a degeneracy between planet radius and the mean geometric albedo. If multi-epoch optical observations are paired with NIR spectra, one could overcome this. A measure of radius, surface gravity and effective temperature can be obtained using the NIR spectrum and system age, as has long been done for known self-luminous planets. This will set the radius and the level of phase-independent thermal emission in the optical. Then, one could use the multi-epoch optical measurements to empirically back out $\Phi(\alpha)$ and $A_g$ without the usual radius - albedo degeneracy. On the other hand, in the event that one has only a single epoch of optical observations, lack of knowledge of the phase curve and radius could inhibit precise measurements of albedo, metallicity, and sensitivity to cloudiness. 

\begin{figure}[htb!]
\includegraphics[width=0.5\textwidth]{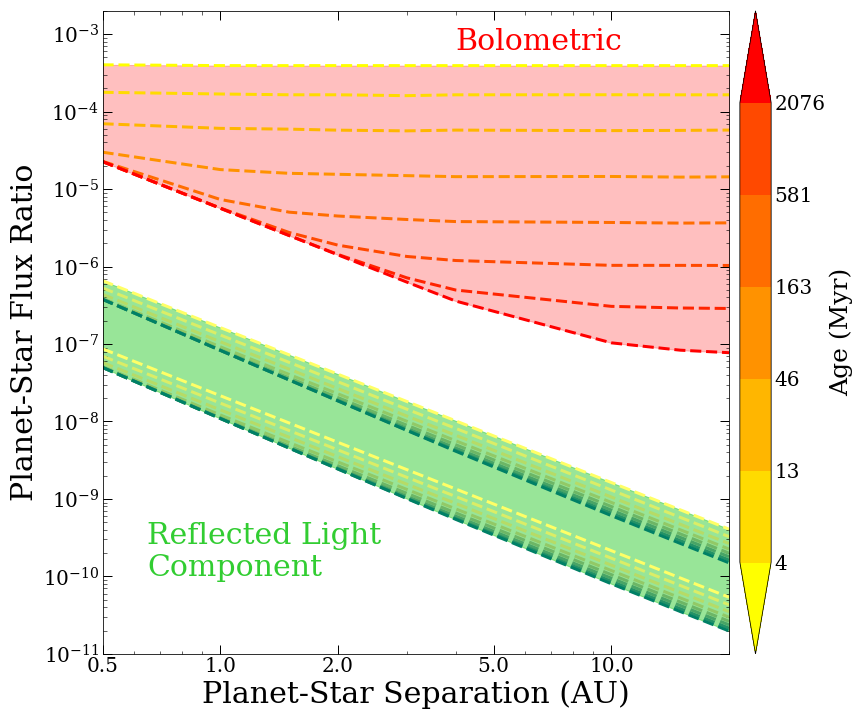}
\includegraphics[width=0.5\textwidth]{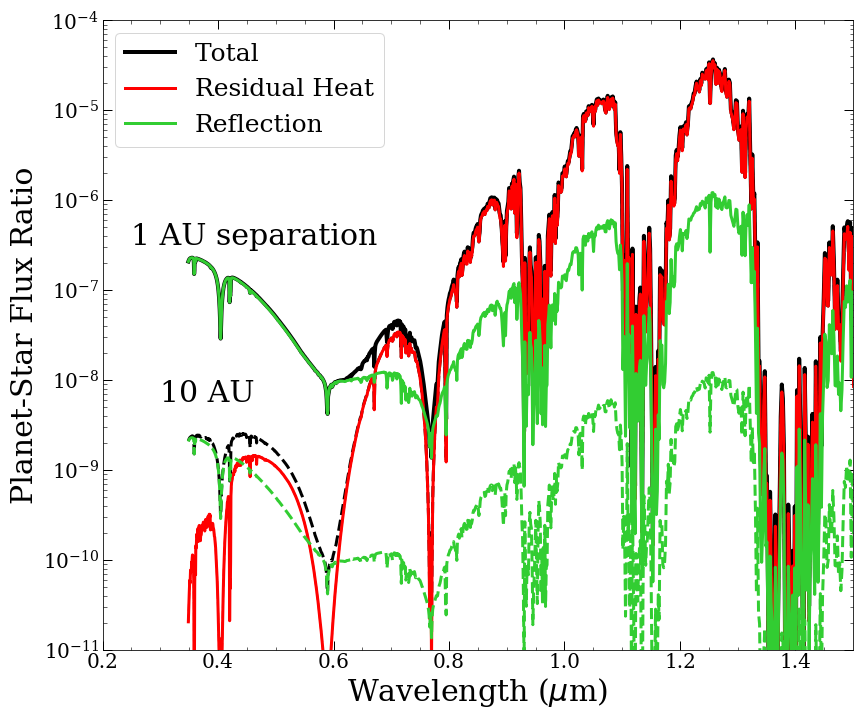}
\caption{\label{reflight_v_residualheat}The \textbf{left} panel shows a visualization of the contribution of reflected light to the bolometric planet-star flux ratio as a function of time and separation for a 1 Jupiter-mass object. The reflected light component varies with radius and separation, bounded on top by the high-albedo full-illumination case and on bottom by the low-albedo half-illumination case. In reality, it is expected that the geometric albedo will exhibit a dependence on temperature and irradiation levels, not accounted for here \citep{sudarsky2000}. The \textbf{right} panel shows the total planet-star flux ratio separated into a contribution from reflected light and thermal emission for 4.64 Myr. This demonstrates that, while the reflected light component remains over an order of magnitude less than the bolometric flux ratio across all the ages and separations modeled here, at shorter wavelengths, reflected light can still dominate.}
\end{figure}

The left panel of Fig. \ref{reflight_v_residualheat} shows the bolometric planet-star flux ratio and the reflected light planet-star flux ratio as a function of time and planet-star separation for a 1-M$_J$ planet. The upper bound for reflected light corresponds to the planet as seen at full illumination with an albedo of 0.5, similar to that expected for Jupiter, while the lower bound corresponds to the planet as seen at a half illuminated crescent phase with an albedo of 0.1 as has been measured for hot Jupiters. We assume a Lambertian phase function in computing the half-illuminated planet-star flux ratio because it is convenient and qualitatively reasonable. This gives a rough sense of the degree to which reflected light will contribute to the spectra of planets within 20 AU. The right panel of Fig. \ref{reflight_v_residualheat} shows an approximate decomposition of our model spectra into the reflected light and residual heat components. While the reflected light flux never reaches the bolometric flux, at wavelengths $<$0.6 $\mu$m it can still be the dominant contributor to the planet's spectrum if the orbital separation is at 1 AU. Our code does not do a photon-by-photon calculation as one might with a Monte Carlo radiative transfer code, so it is impossible to assign definitive origins to all of the flux in the output spectra. However, we can utilize our knowledge of reflected light planet-star flux ratios along with some reasonable assumptions in order to disentangle the reflection and thermal emission after-the-fact. The main assumption we make is that our planets at 10 and 1 AU share an identical thermal emission component to their spectrum. Looking at Table \ref{hyp_tab}, one can see that, at earlier ages, values of surface gravity, effective temperature and radius are very similar or identical, so differences in thermal emission can only arise from a variation in the shape of the temperature-pressure profile induced by varying levels of stellar irradiation. 

First, we break our total planet star flux ratio into the sum of a thermal planet-star flux ratio and a reflected light planet-star flux ratio.
\begin{equation}
    \left(\frac{F_{P}}{F_{*}}\right)_{tot} = \left(\frac{F_{P}}{F_{*}}\right)_{ref} + \left(\frac{F_{P}}{F_{*}}\right)_{therm} \rm.
\end{equation}

Taking equation (2) for a planet at 1 AU and its counterpart at 10 AU and subtracting the two, we get the following:
\begin{equation}
    \left(\frac{F_{P}}{F_{*}}\right)_{tot, 1 AU} - \left(\frac{F_{P}}{F_{*}}\right)_{tot, 10 AU} = \left(\frac{F_{P}}{F_{*}}\right)_{ref, 1 AU} - \left(\frac{F_{P}}{F_{*}}\right)_{ref, 10 AU}    \rm.
\end{equation}
If we plug in our expression for the reflected light planet-star flux ratio, we have:
\begin{equation}
    \left(\frac{F_{P}}{F_{*}}\right)_{tot, 1 AU} - \left(\frac{F_{P}}{F_{*}}\right)_{tot, 10 AU} = \Phi(\lambda,\alpha)A_{g, 1 AU}(\lambda)\left(\frac{R_{P, 1 AU}}{1 AU}\right)^2 - \Phi(\lambda,\alpha)A_{g, 10 AU}(\lambda)\left(\frac{R_{P, 10 AU}}{10 AU}\right)^2 \rm.
\end{equation}
In computing our spectra, they are always assumed to be at full illumination, where $\Phi$=1.0 . If we make the additional assumption that the geometric albedos of our planets are the same, then we get the following equation:
\begin{equation}
    \left(\frac{F_{P}}{F_{*}}\right)_{tot, 1 AU} - \left(\frac{F_{P}}{F_{*}}\right)_{tot, 10 AU} = A_{g}(\lambda)\left[\left(\frac{R_{P, 1 AU}}{1 AU}\right)^2 - \left(\frac{R_{P, 10 AU}}{10 AU}\right)^2\right] \rm.
\end{equation}
This can be re-arranged to solve for $A_{g}$. Once we have $A_{g}$, we can use equation (1) subtract the reflected planet-star flux ratio from the total planet star flux ratio and solve for the thermal component alone. Looking at the results of this exercise in the right panel of Fig. \ref{reflight_v_residualheat}, one can see that, consistent with previous studies, reflected light tends to dominate the shape of the spectra below 0.6 $\mu m$, both reflected light and residual heat contribute between 0.6 $\mu$m and 0.8 $\mu$m, and longer than $\sim$0.8 $\mu$m the residual heat dominates.

\begin{figure}[htb!]
\begin{center}
\includegraphics[width=0.85\textwidth]{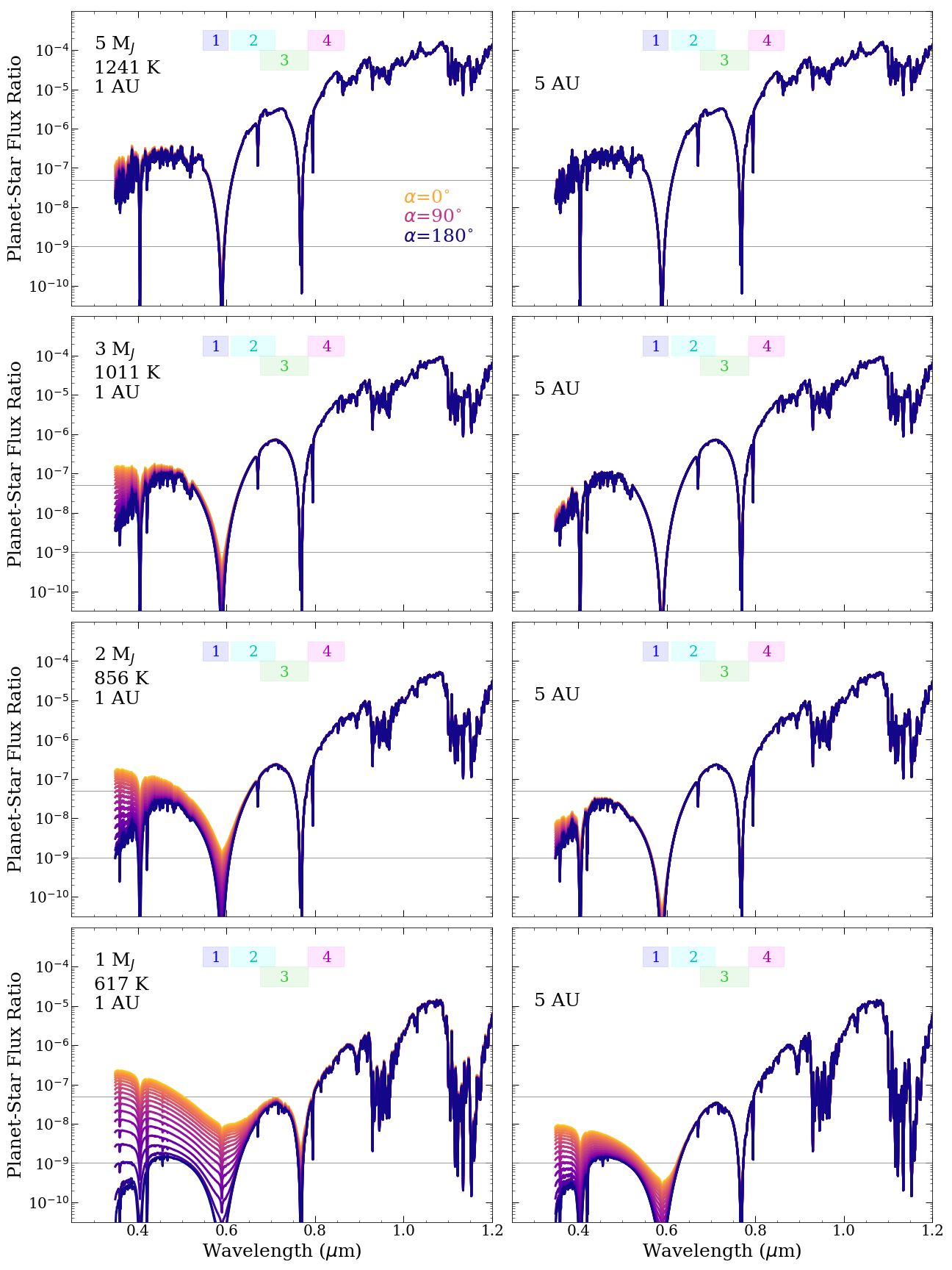}
\end{center}
\caption{\label{phase_dependence_younger}Phase-dependent planet-star flux ratios for a variety of planet masses and separations, all at 4.64 Myr. The \textbf{left} and \textbf{right} columns have planets at 1 and 5 AU, respectively. The rows from top to bottom have 5, 3, 2, and 1 M$_J$. We assume a circular, edge-on system with a Lambertian phase function, and use the procedure for estimating the phase dependence explained in \S \ref{sec:disentangle}. The dark purple color indicates a phase angle of 180$^{\circ}$, when the day side of the planet faces directly away from the observer. The gold color indicates a phase angle of 0$^{\circ}$, when the planet appears fully illuminated. Lines in between step evenly in phase angle. WFIRST-CGI's bandpasses are noted by colored rectangles, and its contrast levels are marked with horizontal black lines. Portions of the spectra where colored lines separate have observable contributions from reflected light.}
\end{figure}

\begin{figure}[htb!]
\begin{center}
\includegraphics[width=0.85\textwidth]{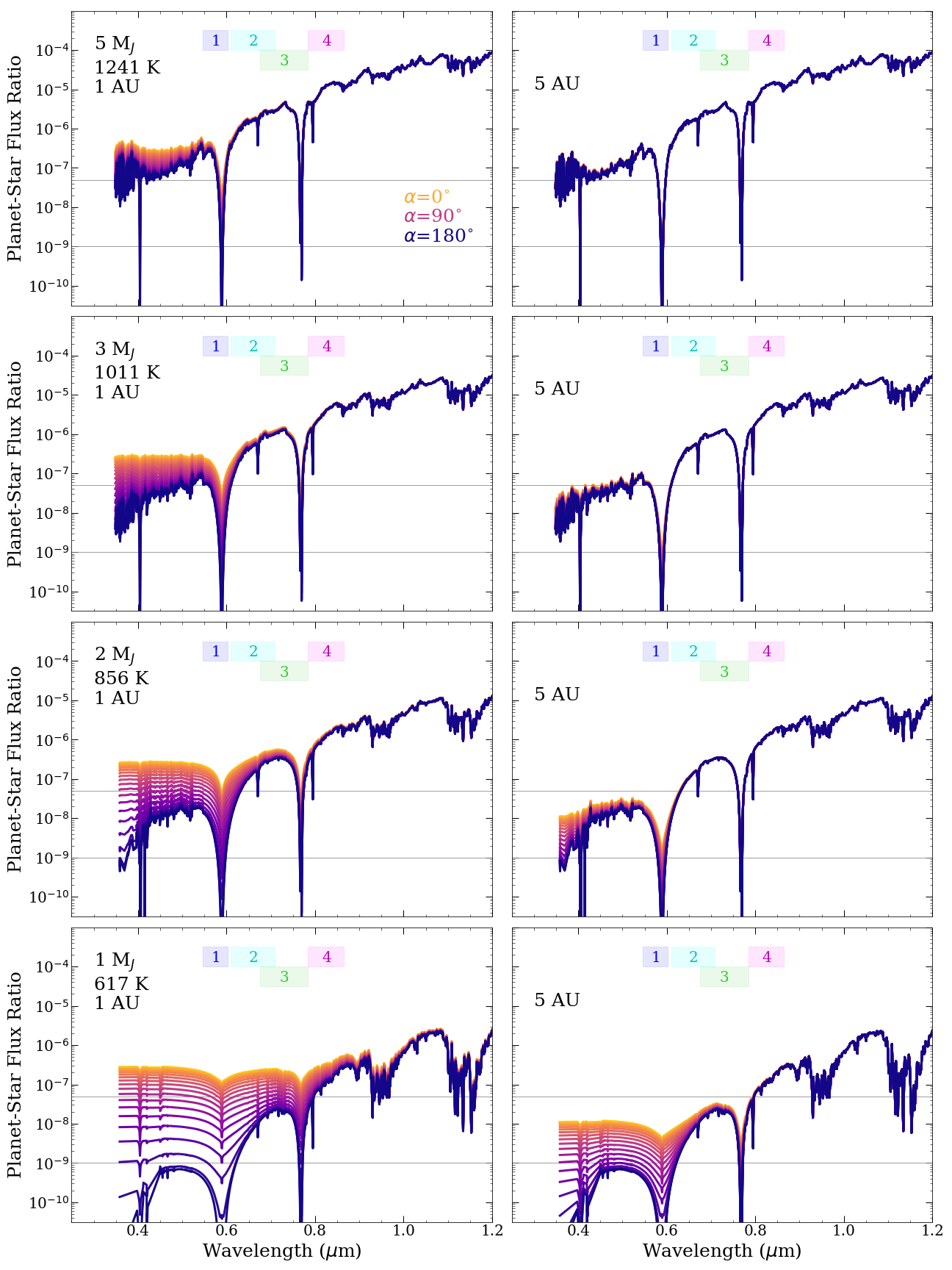}
\end{center}
\caption{\label{phase_dependence_cloudy}Similar to Fig. \ref{phase_dependence_younger}, now showing phase-dependent planet-star flux ratios for the same planets with a scattering gray cloud added through out the whole atmosphere. The \textbf{left} and \textbf{right} columns again have planets at 1 and 5 AU respectively, and all temperatures, radii, and surface gravities are for 4.64 Myr ages. Planets with highly scattering condensates show a contribution from reflected light out to longer wavelengths than their clear counterparts.}
\end{figure}

Figure \ref{reflight_v_residualheat} shows the spectra at full illumination, but, as discussed earlier, there will be a phase dependence of young giant planet optical spectra. To examine this, we assume that the thermal component of the planet-star flux ratio remains constant throughout the orbit, while the reflected light component changes with phase according to a Lambertian phase function. This yields the results shown in Fig. \ref{phase_dependence_younger} and Fig. \ref{phase_dependence_cloudy} for planets with clear and cloudy atmospheres, respectively. This procedure illustrates nicely under which conditions and at what wavelengths reflected light significantly effects the spectra in Fig. \ref{hypothetical_planets}. Anywhere that the plots in Fig. \ref{phase_dependence_younger} do not appear purely dark purple, one can infer that reflected light is noticeably altering the observed spectrum. Because of the steep wavelength dependence of Rayleigh scattering and the deep sodium feature making observations around 0.59 $\mu$m extremely difficult, measurements bluer than 0.5 $\mu$m will have a much better chance of detecting reflected light for clear atmospheres. This is seen clearly in Fig. \ref{phase_dependence_younger}. In Fig. \ref{phase_dependence_cloudy} with an additional gray scattering opacity source, the strongest scattering source does not depend on wavelength, so we see that the variation with phase is strong for wavelengths all the way to 0.7 $\mu$m. 

Our procedure made some strong assumptions: first, that planets with the same temperatures and surface gravities, but differing levels of stellar irradiation will have the same thermal emission component, and, second, that A$_{g}$ for our planets at 1 and 10 AU are identical. Given this, the phase-dependent spectra should be taken only as rough estimates made to illustrate a key point: multi-epoch observations made in the optical and paired with NIR spectra break the degeneracy between radius and geometric albedo and enable an empirical measurement of the planet's phase function. Future missions with wavelength coverage extending blueward of 0.6 $\mu$m, and smaller inner working angles will make this valuable type of observation quite feasible (see Fig. \ref{fig:tongue}). For most of the self-luminous planets and substellar companions likely to be observed with WFIRST-CGI, reflected light will not be contributing because they orbit at separations greater than 10 AU and have temperatures over 1000 K. For 51 Eri b, which has a temperature of only $\sim$700 K \citep{eri-b} and an eccentric orbit, the variation of reflected light with phase could be relevant. This is especially relevant if its atmosphere contains highly scattering condensates, but at a temperature of 700 K it is not known what species may be present.

\section{Summary and Conclusions}\label{sec:conclusion}

In this paper, we have laid out a case for why young giant planets could be a feasible and scientifically valuable target for future direct-imaging missions with optical wavelength coverage:  

\textbf{(1) They will be observable in the optical, and offer an easier intermediate target compared with mature giant planets and exoearths.} Future space-based direct-imaging missions will achieve higher contrast ratios at bluer wavelengths as they aim to image exoearths. Such instruments should easily attain high-precision measurements of young self-luminous planets in a novel wavelength range. The technology demonstration WFIRST-CGI mission will already be able to obtain spectra for a handful of these objects, providing further constraints on the cloudiness and metallicity of a few imaged self-luminous planets and substellar companions (e.g., HD 206893B, HD 984B, and HR 8799e). HST-STIS may be able to observe some known objects as well. A future LUVOIR type mission will be able to observe even more. 

\textbf{(2) It is expected that a sizeable population of young-adolescent planets at moderate separations exists to be discovered and observed with the next generation of direct-imaging technology.} Based upon RV studies and the first generation of direct-imaging surveys, we have learned that the occurrence rate of Jupiter-sized planets peaks between 3-10 AU. With the advent of ELTs and optical space-based direct imaging missions, these planets should become accessible. Given the population of moving group members younger than 100 Myr within 150 pc, and the occurrence rates for giant planets (around 3.5\% or so) self-luminous, moderate separation planets should be observable.

\textbf{(3) The optical wavelength range is rich with information that can help to constrain the properties of these young giant planets.} Most notably, it contains the strong potassium and sodium resonance doublets. The shape and depth of these features are sensitive to metallicity, surface gravity, temperature, and the nature of any condensates present in the atmosphere. NIR observations of some young giant planets hint at the presence of dust and condensates, but the exact nature of these particulates remains poorly understood. If forsterite clouds are present, then there may be a boost in brightness in the optical range by factors of around 2-4 and a narrowing of the potassium and sodium features. The optical wavelength range could also contain emission features indicative of ongoing accretion in protoplanetary objects. Combining space-based optical measurements with NIR measurements from ELTs will provide the best possible characterizations of such young giant planets. 

\textbf{(4) Planets which exhibit both a significant flux from reflected light and from their residual heat of formation present an opportunity to break the degeneracy between radius and geometric albedo, and to empirically measure their reflected light phase function.} Reflected light phase functions are not well understood, presenting a sizeable uncertainty in yield predictions and signal-to-noise ratio calculations for missions seeking to image light reflected from mature planets. They also provide a constraint on condensate properties. Unfortunately, the known systems which will be observable with WFIRST-CGI are all too far out and too young (i.e. HOT) for reflected light to contribute noticeably to their spectra. In a clear atmosphere with Rayleigh scattering from H$_2$ and He as the main source of scattering; when planets cool below ~900 K, reflection begins to play a role at wavelengths shorter than $\sim$0.8 $\mu m$ for separations of 1 AU. At larger separations of around 10 AU, planets need to cool to around 700 K for reflected light to contribute significantly. 51 Eri b has a highly eccentric orbit and approaches within 6 AU of its host star. This is right on the boundary of enabling such measurements. Comparing the albedos and phase functions of young self-luminous planets to those of hot Jupiters would be an interesting exercise in comparative planetology since both populations occupy similar ranges of effective temperature.

The authors hope that this brief exploration will inspire the direct imaging community to include optical observations of young giant planets in their mission planning, and to further consider what additional insights may be gleaned from this class of objects.     

\acknowledgments
The authors would like to acknowledge support for this research under NASA 
WFIRST-SIT award \# NNG16PJ24C and NASA Grant NNX15AE19G, and discussions with the WFIRST-CGI SIT. This research has made use of the NASA Exoplanet Archive, which is operated by the California Institute of Technology, under contract with the National Aeronautics and Space Administration under the Exoplanet Exploration Program.

\vspace{5mm}

\software{astropy \citep{astropy2013}}


\end{document}